%% file: article.tex
\documentclass[journal=jctcce,manuscript=article,layout=twocolumn]{achemso}

\SectionNumbersOn

\usepackage[svgnames]{xcolor}
\usepackage[version=3]{mhchem}
\usepackage{subcaption}
\usepackage{physics}
\usepackage{pgfplots}
\usepackage{bm}
\pgfplotsset{compat=1.18}
\usetikzlibrary{positioning, decorations.pathreplacing, calc, intersections, pgfplots.groupplots}
\usepackage{booktabs}

\definecolor{color1}{HTML}{468B97}
\definecolor{color2}{HTML}{8EAC50}
\definecolor{color3}{HTML}{EF6262}
 
\definecolor{crimson2143940}{RGB}{214,39,40}
\definecolor{darkgray176}{RGB}{176,176,176}
\definecolor{darkorange25512714}{RGB}{255,127,14}
\definecolor{darkturquoise23190207}{RGB}{23,190,207}
\definecolor{forestgreen4416044}{RGB}{44,160,44}
\definecolor{goldenrod18818934}{RGB}{188,189,34}
\definecolor{gray127}{RGB}{127,127,127}
\definecolor{mediumpurple148103189}{RGB}{148,103,189}
\definecolor{orchid227119194}{RGB}{227,119,194}
\definecolor{sienna1408675}{RGB}{140,86,75}
\definecolor{steelblue31119180}{RGB}{31,119,180}
\definecolor{orange1}{HTML}{F2BE22}
\definecolor{orange2}{HTML}{F24C3D}
\definecolor{orange3}{HTML}{FFA500}

\pgfplotscreateplotcyclelist{matplotlib}{
steelblue31119180\\
darkorange25512714\\
forestgreen4416044\\
crimson2143940\\
mediumpurple148103189\\
sienna1408675\\
orchid227119194\\
gray127\\
goldenrod18818934\\
darkturquoise23190207\\
steelblue31119180\\
}

\author{Juan Felipe Huan Lew-Yee}
\email{felipe.lew.yee@quimica.unam.mx}
\affiliation{Departamento de F\'isica y Qu\'imica Te\'orica, Facultad de Qu\'imica,
Universidad Nacional Aut\'onoma de M\'exico, M\'exico City, C.P. 04510,
M\'exico}

\author{Iván Alejandro Bonfil-Rivera}
\affiliation{Departamento de F\'isica y Qu\'imica Te\'orica, Facultad de Qu\'imica,
Universidad Nacional Aut\'onoma de M\'exico, M\'exico City, C.P. 04510,
M\'exico}

\author{Mario Piris}
\email{mario.piris@ehu.eus}
\affiliation{Donostia International Physics Center (DIPC) and Kimika Fakultatea, Euskal Herriko Unibertsitatea (UPV/EHU), 20018 Donostia, Spain.}
\alsoaffiliation{IKERBASQUE, Basque Foundation for Science, 48013 Bilbao, Spain.}

\author{Jorge M. del Campo}
\email{jmdelc@unam.mx}
\affiliation{Departamento de F\'isica y Qu\'imica Te\'orica, Facultad de Qu\'imica,
Universidad Nacional Aut\'onoma de M\'exico, M\'exico City, C.P. 04510,
M\'exico}

\title {Excited states by coupling Piris natural orbital functionals with the extended random phase approximation}

\abbreviations{PNOF,ERPA}

\begin{document}

\begin{abstract}
In this work, we explore the use of Piris natural orbital functionals (PNOFs) to calculate excited-state energies by coupling their reconstructed second-order reduced density matrix with the extended random-phase approximation (ERPA). We have named the general method as PNOF-ERPA, and specific approaches are referred to as PNOF-ERPA0, PNOF-ERPA1, and PNOF-ERPA2, according to the way the excitation operator is built. The implementation has been tested in the first excited states of \ce{H2}, \ce{HeH+}, \ce{LiH}, \ce{Li2}, and \ce{N2}, showing good results compared to the configuration interaction (CI) method. As expected, an increase in accuracy is observed when going from ERPA0 to ERPA1 and ERPA2.  We have also studied the effect of electron correlation included by PNOF5, PNOF7, and the recently proposed global NOF (GNOF) on the predicted excited states. PNOF5 appears to be good and may even provide better results in very small systems, but including more electron correlation becomes important as the system size increases, where GNOF achieves better results. Overall, the extension of PNOF to excited states has been successful, making it a promising method for further applications.
\end{abstract}

\section{Introduction}

Excited states\cite{Adamson1983-fi,Matsika2018-zs} are important for the description of photochemical\cite{Demchenko2017-yq} and electrochemical processes,\cite{Choi2021-mk} fluorescence and phosphorescence phenomena,\cite{Itoh2012-iz} spectroscopy,\cite{Norman2018-vw} and chemical reaction mechanisms,\cite{Su2008-fg,Yu2011-hz} with a variety of cutting-edge chemical applications such as the development of new materials for organic solar cells\cite{Hustings2022-tx} and batteries.\cite{Wang2020-gn,Mandal2022-oa,Fasulo2023-bu} The energy of these states can be calculated using the configuration interaction (CI) method; however, this becomes too expensive even for low levels of CI, although some variations have been developed to address this issue.\cite{Kossoski2023-ve}. There are other electronic structure tools that allow studying excited species at a more affordable cost, such as the equation-of-motion coupled-cluster (EOM-CC)\cite{Emrich1981,Gwaltney1996,Krylov2008} approach, the time-dependent density functional theory (TD-DFT),\cite{Runge1984} and the random phase approximation\cite{Ren2012a}, but the accuracy achieved is not as good as desired and the picture becomes more complex when it comes to static correlation, since multireference methods are required.\cite{Lischka2018-yz,Loos2020-hd}

In this context, the one-particle reduced density matrix (1RDM) functional theory (1RDMFT)\cite{Gilbert1975,Valone1980} appears as a suitable approach to study excited states taking into account electronic correlation effects, including the strong ones. In particular, time-dependent 1RDMFT in its adiabatic linear response formulation has been developed\cite{Pernal2007a,Pernal2007} to calculate the energies of excited states and oscillator strengths,\cite{van-Meer2013-bk} however, a solid foundation for a dynamic 1RDMFT is still an open challenge.\cite{Pernal2016} On the other hand, an ensemble version of 1RDMFT has recently been proposed\cite{Schilling2021} to calculate the energies of selected low-lying excited states, although it will require more efficient numerical minimization schemes for its future success.\cite{Liebert2022} In this article, we shall use the extended random phase approximation\cite{Gambacurta2008,Gambacurta2009} within the 1RDMFT framework in the natural orbital representation. Specifically, we will employ the Chatterjee and Pernal's formulation\cite{Chatterjee2012-qn} that relies on the 1RDM and the two-particle reduced density matrix (2RDM) of the ground state. The method can be elegantly derived from the formally exact Rowe's excitation operator equation-of-motion\cite{Rowe1968-yw}, and has been used\cite{Pernal2014-he,Pernal2014-zb,Pastorczak2015-wz} successfully with the RDMs corresponding to the wavefunction of the antisymmetrized product of strongly orthogonal geminals (APSG).

In this vein, it has been shown\cite{Pernal2013-if,Piris2013-qj} that the APSG approach is equivalent to the Piris natural orbital functional 5 (PNOF5),\cite{Piris2011-qn} except for a phase factor. PNOFs\cite{Piris2006-kn,Piris2013b} are based on the reconstruction of the 2RDM constrained to certain bounds due to the N-representabilty conditions\cite{Piris2018d}, and belong to the JKL-only family of natural orbital functional (NOFs), where J and K refer to the usual Coulomb and exchange integrals, while L denotes the exchange-time-inversion integral.\cite{Piris1999} The latter is relevant for excited states due to the fact that it allows the time-evolution of the occupation numbers, contrary to the stationarity of the occupation numbers demonstrated\cite{Pernal2007a,Benavides-Riveros2019} for the JK-only NOFs. 

The performance of PNOFs has achieved chemical accuracy in many cases,\cite{Lopez2010} with electron-pairing based functionals\cite{Piris2018a} being particularly successful in describing nondynamic electron correlation, namely PNOF5,\cite{Piris2013-qj} PNOF6,\cite{Piris2014c} and PNOF7.\cite{Piris2017-go} Furthermore, the most recent functional, GNOF\cite{Piris2021-sv} has extended the success of PNOF to a balanced electron correlation regime,\cite{Lew-Yee2023-vf} as has been observed in the study of a variety of chemical systems such as hydrogen models in one, two, and three dimensions,\cite{Mitxelena2022-fs} iron porphyrin multiplicity,\cite{Lew-Yee2023-ke} carbenes singlet-triplet gaps,\cite{Franco2023-yf} and all-metal aromaticity\cite{Mercero2023-mw}. Motivated by this success, the extension of the PNOF to excited states becomes tempting, which can be accomplished by introducing their approximate RDMs into the ERPA equations. The PNOF-ERPA approach has the potential to be a viable substitute for multireference wavefunction techniques in the modeling of excited states.

This article is organized as follows. In Section~\ref{sec:Theory}, we briefly review the equations of the ERPA and the PNOFs used. Next, we give some computational details of the calculations in Section~\ref{sec:Computational}. In Section~\ref{sec:Results}, the performance of these approaches is tested in detail on potential energy curves (PECs) of diatomic molecules with increasing number of electrons. Finally, conclusions are provided in Section~\ref{sec:Conclusions}.

\section{\label{sec:Theory}Theory}

In this section, we summarize the ERPA equations as used in this work to couple them with the reconstructed 2RDM of PNOFs in terms of occupation numbers. A detailed description of ERPA can be found in the work of Chatterjee and Pernal.\cite{Chatterjee2012-qn} We address only singlet states, so we adopt a spin-restricted formalism in which a single set of orbitals is used for alpha and beta spins.

\subsection{ERPA}

In the context of the equation-of-motion method, the expectation value of the double commutator developed by Rowe\cite{Rowe1968-yw} for a system described by a Hamiltonian $\hat{H}$ is defined as

\begin{multline}
    \bra{\psi_0} \comm{\delta \hat{O}_\nu}{\comm{\hat{H}}{\hat{O}_\nu^\dagger}} \ket{\psi_0}\\ = \omega \bra{\psi_0} \comm{\delta \hat{O}_\nu}{\hat{O}_\nu^\dagger} \ket{\psi_0} \>\>,
    \label{eq:eom}
\end{multline}
where $\omega$ corresponds to the excitation energy, $\hat{O}_\nu^\dagger$ is the excitation operator that applied to the ground state $\ket{\psi_0}$ produces the excited state $\ket{\psi_\nu}$, namely
\begin{equation}
    \hat{O}^\dagger_\nu \ket{\psi_0} = \ket{\psi_\nu} \>\>,
\end{equation}
whereas $\hat{O}_\nu$ deexcitates from $\ket{\psi_\nu}$ to $\ket{\psi_0}$, and satisfies the consistency condition to ensure the orthogonality of the ground and excited states, that is,
\begin{equation}
    \hat{O}_\nu \ket{\psi_0} = 0 \>\>.
    \label{eq:consistent-condition}
\end{equation}
The equations to solve are obtained by using an excitation operator, with its simplest form including only single non-diagonal excitations, which we have called ERPA0. Therefore, $\hat{O}^\dagger_\nu$ is approximated as 
\begin{eqnarray}
    \hat{O}^\dagger_\nu = &\sum_{p>q}^{N/2} & X_{pq} \left( a_{p_{\alpha}}^\dagger a_{q_{\alpha}} + a_{p_{\beta}}^\dagger a_{q_{\beta}} \right) \nonumber\\
    + &\sum_{p>q}^{N/2} & Y_{pq} \left( a_{q_{\alpha}}^\dagger a_{p_{\alpha}} + a_{q_{\beta}}^\dagger a_{p_{\beta}} \right) \>\>,
    \label{eq:exop0}
\end{eqnarray}
where $X_{pq}$, and $Y_{pq}$ are coefficients to be determined. In the following, the indices $p$, $q$, $r$, $s$, $t$, $u$, and $v$ will be used for spatial orbitals, and $\alpha$ and $\beta$ for spin.

Taking the variation of the adjoint of the excitation operator and substituting in Eq.~(\ref{eq:eom}), it is obtained
\begin{multline}   
2\sum_{r>s}^{N/2} \delta X_{rs} \bigg(\sum_{p>q}^{N/2} X_{pq} A_{srqp} + \sum_{p>q}^{N/2} Y_{pq} A_{srpq}\bigg) \\
+ 2\sum_{r>s}^{N/2} \delta Y_{rs} \bigg(\sum_{p>q}^{N/2} X_{pq} A_{rsqp} + \sum_{p>q}^{N/2} Y_{pq} A_{rspq} \bigg) \\
 = 2\omega \sum_{r>s}^{N/2} \delta X_{rs} \bigg(\sum_{p>q}^{N/2} X_{pq} G_{srqp} + \sum_{p>q}^{N/2} Y_{pq} G_{srpq} \bigg) \\
+ 2\omega \sum_{r>s}^{N/2} \delta Y_{rs} \bigg(\sum_{p>q}^{N/2} X_{pq} G_{rsqp} + \sum_{p>q}^{N/2} Y_{pq} G_{rspq}\bigg) \>\>, 
\label{eq:ERPA0-Complete}
\end{multline}
with
\begin{multline}
    A_{rspq} = \frac{1}{2} \bra{0}[a_{r_{\alpha}}^\dagger a_{s_{\alpha}} + a_{r_{\beta}}^\dagger a_{s_{\beta}},\\ [\hat{H},a_{q_{\alpha}}^\dagger a_{p_{\alpha}} + a_{q_{\beta}}^\dagger a_{p_{\beta}}]]\ket{0} \>\>,
    \label{eq:A_rspq_2nd_quant}
\end{multline}
and
\begin{multline}
    G_{rspq} = \frac{1}{2} \bra{0}[ a_{r_{\alpha}}^\dagger a_{s_{\alpha}} + a_{r_{\beta}}^\dagger a_{s_{\beta}},\\ a_{q_{\alpha}}^\dagger a_{p_{\alpha}} + a_{q_{\beta}}^\dagger a_{p_{\beta}}]\ket{0} \>\>,
    \label{eq:G_rspq_2nd_quant}
\end{multline}
where a factor of ``2'' has been added to Eq.~(\ref{eq:ERPA0-Complete}) and of ``1/2'' to Eqs.~(\ref{eq:A_rspq_2nd_quant}) and~(\ref{eq:G_rspq_2nd_quant}) for convenience.

Considering the 1RDM in its diagonal representation
\begin{equation}
    \Gamma_{rp}^{\sigma_1\sigma_2} = \bra{0} a_{r_{\sigma_1}}^\dagger a_{p_{\sigma_2}} \ket{0} = n_r \delta_{rp} \delta_{\sigma_1 \sigma_2} \>\>,
\end{equation}
the 2RDM
\begin{equation}
    \mathrm{D}_{rspq}^{\sigma_1\sigma_2\sigma_3\sigma_4} = \frac{1}{2} \bra{0} a_{r_{\sigma_1}}^\dagger a_{s_{\sigma_2}}^\dagger a_{q_{\sigma_4}} a_{p_{\sigma_3}} \ket{0} \>\>,
\end{equation}
and recalling the restricted-spin formalism ($n_p = n_p^\alpha = n_p^\beta$, $\phi_p=\phi_p^\alpha=\phi_p^\beta$, and consequently $D^{\alpha\alpha\alpha\alpha} = D^{\beta\beta\beta\beta}$, $D^{\alpha\beta\alpha\beta} = D^{\beta\alpha\beta\alpha}$), the elements of $A_{rspq}$ can be expressed as
\begin{multline}
    A_{rspq} = h_{sq} \delta_{pr} (n_p - n_s) + h_{pr} \delta_{sq} (n_q - n_r)\\
    + \sum_{tu} (\braket{qt}{su} - \braket{qt}{us}) \mathrm{D}_{ptru}^{\alpha\alpha\alpha\alpha}\\
    + \sum_{tu} (\braket{pt}{ru} - \braket{pt}{ur}) \mathrm{D}_{qtsu}^{\alpha\alpha\alpha\alpha}\\
    + \sum_{tu} \braket{qt}{su} \mathrm{D}_{ptru}^{\alpha\beta\alpha\beta} + \sum_{tu} \braket{qt}{us} \mathrm{D}_{ptur}^{\alpha\beta\alpha\beta}\\
    + \sum_{tu} \braket{pt}{ru} \mathrm{D}_{qtsu}^{\alpha\beta\alpha\beta} + \sum_{tu} \braket{pt}{ur} \mathrm{D}_{qtus}^{\alpha\beta\alpha\beta}\\
    + \sum_{tu} \braket{ps}{ut}\mathrm{D}_{qrtu}^{\alpha\alpha\alpha\alpha} - \sum_{tu} \braket{ps}{tu}\mathrm{D}_{qrtu}^{\alpha\beta\alpha\beta}\\
    + \sum_{tu} \braket{qr}{ut}\mathrm{D}_{pstu}^{\alpha\alpha\alpha\alpha} - \sum_{tu} \braket{qr}{tu}\mathrm{D}_{pstu}^{\alpha\beta\alpha\beta}\\
    + \delta_{qs} \sum_{tuv} \braket{pt}{vu}\mathrm{D}_{rtuv}^{\alpha\alpha\alpha\alpha} - \delta_{qs} \sum_{tuv} \braket{pt}{uv}\mathrm{D}_{rtuv}^{\alpha\beta\alpha\beta}\\
    + \delta_{pr} \sum_{tuv} \braket{qt}{vu}\mathrm{D}_{stuv}^{\alpha\alpha\alpha\alpha} - \delta_{pr} \sum_{tuv} \braket{qt}{uv}\mathrm{D}_{stuv}^{\alpha\beta\alpha\beta} \>\>,
    \label{eq:A}
\end{multline}
where $h_{pq}$ represent the elements of the core Hamiltonian matrix, and $\braket{pq}{rs}$ corresponds to the electron repulsion integrals in the basis of spatial natural orbitals. Applying the commutator and considering the 1RDM in its diagonal representation of natural orbitals and occupation numbers, the elements of $G_{rspq}$ are given by
\begin{equation}
    G_{rspq} = \delta_{sq} \delta_{rp} \left(n_r - n_s\right) \>\>.
    \label{eq:G_rspq}    
\end{equation}
Grouping the terms of equation~(\ref{eq:ERPA0-Complete}) by the variations of $\delta X_{rs}$ and $\delta Y_{rs}$ leads to two type of equations. Furthermore, these can be simplified by considering the Kronecker deltas of Eq.~(\ref{eq:G_rspq}) and the conditions $r>s$ and $p>q$ imposed by the sums to give
\begin{multline}
    \forall_{r>s} \sum_{p>q}^{N/2} X_{pq} A_{rspq} + \sum_{p>q}^{N/2} Y_{pq} A_{rsqp}\\ = \omega X_{rs} (n_s-n_r) \>\>,
\end{multline}
\begin{multline}
    \forall_{r>s} \sum_{p>q}^{N/2} X_{pq} A_{rsqp} + \sum_{p>q}^{N/2} Y_{pq} A_{rspq}\\
 = -\omega Y_{rs} (n_s - n_r) \>\>,
\end{multline}
where we have used the fact that
\begin{equation}
    A_{rspq} = A_{srqp} \>\>.
\end{equation}
This equations can be cast in matrix form to the generalized eigenvalue problem given by
\begin{multline}
    \begin{pmatrix}
    A_{rspq} & B_{rspq}\\
    B_{rspq} & A_{rspq}
    \end{pmatrix}
    \begin{pmatrix}
    X_{pq}\\
    Y_{pq}
    \end{pmatrix} \\ \\
    = \omega
    \begin{pmatrix}
    \Delta N_{rspq} & 0\\
    0 & -\Delta N_{rspq}
    \end{pmatrix}
    \begin{pmatrix}
    X_{pq}\\
    Y_{pq} 
    \end{pmatrix} \>\>,
    \label{eq:ERPA0}
\end{multline}
where we have used
\begin{eqnarray}
    B_{rspq} &=& A_{rsqp}\\    
    \Delta N_{rspq} &=& (n_s - n_r) \delta_{rp}\delta_{sq} \>\>.
    \label{eq:DN}
\end{eqnarray}
Eq.~(\ref{eq:ERPA0}) can be written only in terms of $\bf{A}$, but these auxiliary variables allow identifying the appropriate blocks to reformulate the problem in a more compact form, as

\begin{multline}    
    (\bf{A}-\bf{B}) (\bf{\Delta N})^{-1} (\bf{A}+\bf{B}) (\bf{X}+\bf{Y})\\ = \omega^2 \bf{\Delta N} (\bf{X}+\bf{Y}) \>\>,
    \label{eq:reduced-ERPA0-1}
\end{multline}
\begin{multline}    
    (\bf{A}+\bf{B}) (\bf{\Delta N})^{-1} (\bf{A}-\bf{B}) (\bf{X}-\bf{Y})\\ = \omega^2 \bf{\Delta N} (\bf{X}-\bf{Y}) \>\>,
    \label{eq:reduced-ERPA0-2}
\end{multline}

which resemble what is done to reduce the generalized eigenvalue problem of TD-SCF.\cite{Stratmann1998-ox}

Recalling the fact that when having occupation numbers of exactly ones and zeros the PNOFs ground state goes to the Hartree-Fock limit, it is also interesting that in this limit PNOF-ERPA0 becomes equivalent to the TD-HF method. This can be seen from Eq.~(\ref{eq:DN}), where having only ones and zeros as occupation numbers makes $\Delta N$ the identity matrix with some additional zeros that can be discarded. Hence, Eq.~(\ref{eq:ERPA0}) introduces electron correlation to excited states through the occupation numbers.

We can go beyond by including single-diagonal excitations in the operator, a procedure that we have labeled as ERPA1. For this case, the excitation operator is defined as
\begin{eqnarray}
    \hat{O}^\dagger_\nu = &\sum_{p>q}^{N/2} & X_{pq} \left( a_{p_{\alpha}}^\dagger a_{q_{\alpha}} + a_{p_{\beta}}^\dagger a_{q_{\beta}} \right) \nonumber\\
    + &\sum_{p>q}^{N/2} & Y_{pq} \left( a_{q_{\alpha}}^\dagger a_{p_{\alpha}} + a_{q_{\beta}}^\dagger a_{p_{\beta}} \right) \nonumber\\
    + &\sum_{p}^{N/2} & Z_{p} \left( a_{p_{\alpha}}^\dagger a_{p_{\alpha}} + a_{p_{\beta}}^\dagger a_{p_{\beta}} \right) \>\>,
    \label{eq:exop1}
\end{eqnarray}
where $X_{pq}$, $Y_{pq}$ and $Z_{p}$ are the coefficients to be determined. Substituting in Eq.~(\ref{eq:eom}) and following a similar procedure than before, we arrive to
\begin{multline}
    \forall_{r>s} \sum_{p>q}^{N/2} X_{pq} A_{rspq} + \sum_{p>q}^{N/2} Y_{pq} A_{rsqp} + \sum_{p}^{N/2} Z_{p} A_{rspp}\\ = \omega X_{rs} (n_s-n_r) \>\>,
\end{multline}
\begin{multline}
    \forall_{r>s} \sum_{p>q}^{N/2} X_{pq} A_{rsqp} + \sum_{p>q}^{N/2} Y_{pq} A_{rspq} + \sum_{p}^{N/2} Z_{p} A_{rspp}\\
 = -\omega Y_{rs} (n_s - n_r) \>\>,
\end{multline}
\begin{multline}
    \forall_{r} \sum_{p>q}^{N/2} X_{pq} A_{rrpq} + \sum_{p>q}^{N/2} Y_{pq} A_{rrpq} + \sum_{p}^{N/2} Z_{p} A_{rrpp}\\
 = 0 \>\>.
\end{multline}

Defining the auxiliary variables
\begin{eqnarray}
    C_{rsp} &=& A_{rspp}\\    
    E_{rpq} &=& A_{rrpq}\\    
    F_{rp} &=& A_{rrpp} \>\>,
\end{eqnarray}
these equations can be cast in matrix form as
\begin{multline}
    \begin{pmatrix}
    A_{rspq} & B_{rspq} & C_{rsp}\\
    B_{rspq} & A_{rspq} & C_{rsp}\\
    E_{rpq} & E_{rpq} & F_{rp}
    \end{pmatrix}
    \begin{pmatrix}
    X_{pq}\\
    Y_{pq}\\
    Z_{p}
    \end{pmatrix} \\ \\
    = \omega
    \begin{pmatrix}
    \Delta N_{rspq} & 0 & 0\\
    0 & -\Delta N_{rspq} & 0\\
    0 & 0 & 0
    \end{pmatrix}
    \begin{pmatrix}
    X_{pq}\\
    Y_{pq}\\
    Z_{p}
    \end{pmatrix} \>\>.
    \label{eq:ERPA}
\end{multline}
Furthermore, the problem can be reformulated in a more compact form as

\begin{multline}
    (\bf{A}+\bf{B} - 2\bf{C}\bf{F}^{-1} \bf{E}) (\bf{\Delta N})^{-1} (\bf{A}-\bf{B})\\ \times (\bf{X}-\bf{Y}) = \omega^2 \bf{\Delta N} (\bf{X}-\bf{Y}) \>\>,
    \label{eq:reduced-ERPA1-1}
\end{multline}
\begin{multline}
    (\bf{A}-\bf{B}) (\bf{\Delta N})^{-1} (\bf{A}+\bf{B} - 2\bf{C}\bf{F}^{-1} \bf{E})\\ \times (\bf{X}+\bf{Y}) = \omega^2 \bf{\Delta N} (\bf{X}+\bf{Y}) \>\>.
    \label{eq:reduced-ERPA1-2}
\end{multline}

Unfortunately, both ERPA0 and ERPA1 violate the consistency condition, and hence the excitation energies deteriorate. This condition may be enforced for two-electron systems by including double diagonal excitations, namely, ERPA2,\cite{Chatterjee2012-qn} with the excitation operator given by
\begin{eqnarray}
    \hat{O}^\dagger_\nu = &\sum_{p>q}^{N/2}& X_{pq} \left( a_{p_{\alpha}}^\dagger a_{q_{\alpha}} + a_{p_{\beta}}^\dagger a_{q_{\beta}} \right)\nonumber\\ + &\sum_{p>q}^{N/2}& Y_{pq} \left( a_{q_{\alpha}}^\dagger a_{p_{\alpha}} + a_{q_{\beta}}^\dagger a_{p_{\beta}} \right)\nonumber\\ + &\sum_{pq}^{N/2}& W_{pq} \left( a_{p_{\beta}}^\dagger a_{q_{\beta}} a_{p_{\alpha}}^\dagger a_{q_{\alpha}} \right) \>\>.
\end{eqnarray}
Substituting in Eq.~(\ref{eq:eom}), imposing the consistency condition and taking into account that for two-electron systems the RDMs of order higher than two vanish, the equations obtained for two-electron systems are extended to any N-electron spin-compensated system, resulting in the following generalized eigenvalue problem
\begin{multline}
    \begin{pmatrix}
    A_{rspq} & B_{rspq} & \displaystyle \frac{c_s}{2c_p(c_r+c_s)} C_{rsp}\\
    B_{rspq} & A_{rspq} & \displaystyle \frac{c_r}{2c_p(c_r+c_s)} C_{rsp}\\
    \displaystyle \frac{1}{c_r} E_{rpq} & \displaystyle \frac{1}{c_r} E_{rpq} & \displaystyle \frac{1}{4c_pc_r} F_{rp}
    \end{pmatrix}
    \begin{pmatrix}
    X_{pq}\\
    Y_{pq}\\
    V_{p}
    \end{pmatrix} \\ \\=
    \omega
    \begin{pmatrix}
    \Delta N_{rspq} & 0 & 0\\
    0 & -\Delta N_{rspq} & 0\\
    0 & 0 & 1
    \end{pmatrix}
    \begin{pmatrix}
    X_{pq}\\
    Y_{pq}\\
    V_{p} 
    \end{pmatrix} \>\>,
    \label{eq:ERPA2}
\end{multline}
where
\begin{equation}
    V_p = \sum_q W_{pq} c_q \>\>,
\end{equation}
and $c_q$ is the square root of the occupation number $n_q$, that is, $c_q = \pm \sqrt{n_q}$. An important difference between the APSG-ERPA and PNOF-ERPA approaches is the square root sign that is determined in the optimization process of APSG, but it is chosen to reproduce the functional form of PNOF, as will be detailed in the next section.

\subsection{PNOF}

Having the equations of ERPA0, ERPA1 and ERPA2, we only need to express the elements of the matrix $\bf{A}$ given in Eq. (\ref{eq:A}), according to the 2RDM reconstructions of the PNOFs that we consider in this work. This allowed us to implement the PNOF-ERPA0, PNOF-ERPA1 and PNOF-ERPA2 approximations. In the following, we describe the 2RDMs corresponding to PNOF5,\cite{Piris2013-qj} PNOF7\cite{Piris2017-go} and GNOF.\cite{Piris2021-sv}

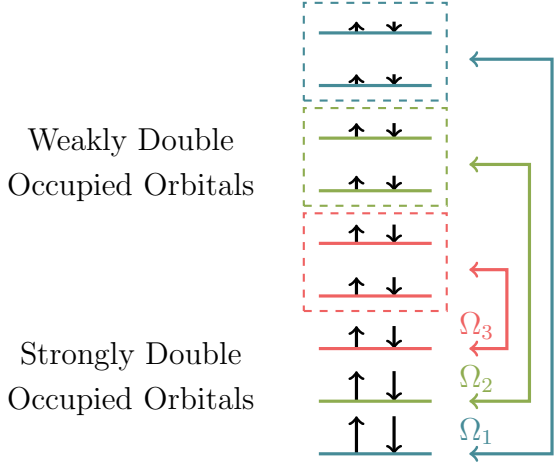
\begin{figure}[htbp]
    \centering
    \input{pairing_scheme}
    \caption{Example of the pairing scheme used in PNOF for a singlet state of a system with 6 electrons ($N=6$). There are $N/2=3$ subspaces, namely, \textcolor{color1}{$\Omega_1$}, \textcolor{color2}{$\Omega_2$} and \textcolor{color3}{$\Omega_3$}. In this example, an extended pairing scheme with $N_g=2$ have been used, therefore there are two weakly double-occupied natural orbitals coupled to each strongly double-occupied natural orbital.}
    \label{fig:pairing-scheme}
\end{figure}

The aforementioned functionals use an electron pairing scheme, as depicted in Fig.~\ref{fig:pairing-scheme}. Given a system with $N$ electrons in the orbital space $\Omega$, we divide the latter into $N/2$ mutually disjoint subspaces $\Omega_g$, so each orbital belongs only to one subspace. A given subspace $\Omega_g$ contains one strongly double-occupied orbital $\phi_g$ below the level $N/2$, and $N_g$ weakly double-occupied orbitals above it, and its occupation numbers sum to ``1''. The case where $N_g=1$ is called the orbital perfect-pairing scheme, while $N_g>1$ is called the extended-pairing scheme. It is important to note that orbitals satisfying the pairing conditions are not required to remain fixed throughout the orbital optimization process. The 2RDM ($\boldsymbol{\mathrm{D}}$) is divided into intra- and inter-subspace contributions, corresponding to the intra-pair electronic correlation, that is, the contribution of the orbitals in a given subspace $\Omega_g$, and the inter-pair electronic correlation, that is, the contribution between the orbitals of a subspace $\Omega_g$ with those of a different subspace $\Omega_f$, $f \neq g$.

The simplest way to meet all N-representability constraints imposed\cite{Piris2006-kn} on the 2RDM of PNOF leads to the independent pairs model PNOF5, where only intra-pair (intra-subspace) electron correlation is taken into account, namely
\begin{equation}
    \mathrm{D}_{pqrt}^{\alpha\beta\alpha\beta} = \frac{\Pi_{pr}}{2} \delta_{pq} \delta_{rt} \delta_{p\Omega_g} \delta_{r\Omega_g} \>\>,
    \label{eq:intra}
\end{equation}
where Kronecker deltas have a standard meaning, for example $\delta_{r\Omega_g}$ is one if the natural orbital $\phi_r$ belongs to the subspace $\Omega_g$, and zero otherwise. The matrix elements are defined as $\Pi_{pr} = c_p c_r$, where $c_p$ is defined by the square root of the occupation numbers according to the rule
\begin{equation}
    c_p = \left.
  \begin{cases}
    \phantom{+}\sqrt{n_p}, & p\leq N/2\\
    -\sqrt{n_p}, & p>N/2 \\
  \end{cases}
  \right. \>\>,
  \label{eq:PNOF5-roots}
\end{equation}
that is, the phase factor of $c_p$ is chosen to be $+1$ for the strongly occupied orbital of a given subspace $\Omega_g$, and $-1$ otherwise. On the other hand, the inter-subspace contributions ($\Omega_g \neq \Omega_f$) are assumed Hartree-Fock-like,
\begin{equation}
    \mathrm{D}_{pqrt}^{\alpha\alpha\alpha\alpha} = \frac{n_p n_q}{2} \left( \delta_{pr} \delta_{qt} - \delta_{pt} \delta_{qr}\right) \delta_{p\Omega_f} \delta_{q\Omega_g} \>\>,
    \label{eq:parallel}
\end{equation}
\begin{equation}
    \mathrm{D}_{pqrt}^{\alpha\beta\alpha\beta} = \frac{n_p n_q}{2} \delta_{pr}\delta_{qt} \delta_{p\Omega_f}\delta_{q\Omega_g} \>\>.
    \label{eq:pnof5}    
\end{equation}

To go beyond the independent-pair approximation, electron correlation between pairs (subspaces) is introduced. In all post-PNOF5 reconstructions, the parallel spin blocks have remained Hartree-Fock-like as in Eq. (\ref{eq:parallel}), while the opposite spin contribution between pairs (subspaces) is different.

For PNOF7, it was introduced the function
\begin{equation}
    \Phi_p = \sqrt{n_p (1-n_p)} \>\>,
\end{equation}
so the interpair opposite spin contribution is given by
\begin{multline}
    \>\>\>\> \mathrm{D}_{pqrt}^{\alpha\beta\alpha\beta} = \frac{1}{2} \bigg( n_p n_q \delta_{pr} \delta_{qt} \delta_{p\Omega_f} \delta_{q\Omega_g} \\
    - \Phi_p \Phi_r \delta_{pq} \delta_{rt} \delta_{p\Omega_f} \delta_{r\Omega_g} \bigg) \>. \>\>\>\>\>\>\>\>
    \label{eq:pnof7}    
\end{multline}    
Note that $\Phi_p$ has significant values only when the occupancies differ substantially from ``1'' and ``0''. Consequently, PNOF7 can recover static correlation between pairs, but it lacks interpair dynamic electron correlation.

\begin{figure*}[htb]
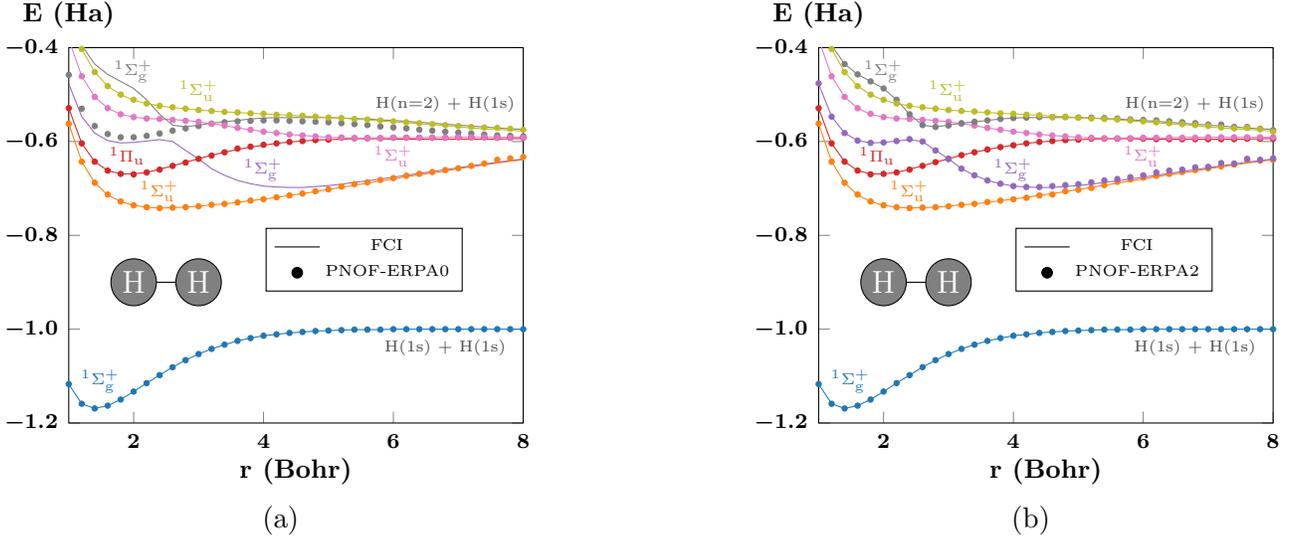

    \centering
    \centering
    \begin{subfigure}[b]{0.43\textwidth}
    \include{H2-ERPA0}
    \vspace{-1.0cm}
    \caption{}
    \label{fig:h2-pnof-erpa0}        
    \end{subfigure}
    \hfill
    \centering
    \begin{subfigure}[b]{0.43\textwidth}
    \include{H2-ERPA2}
    \vspace{-1.0cm}
    \caption{}
    \label{fig:h2-pnof-erpa2}        
    \end{subfigure}
    \hfill
    \caption{PECs of the first states of \ce{H2} computed using a) PNOF-ERPA0 and b) PNOF-ERPA2. FCI results are shown as solid lines as a reference. There are $N_\text{cwo} = 17$  orbitals paired to each strongly double occupied orbital. The first curve corresponds to the ground state.}
    \label{fig:h2-pnof}
\end{figure*}

GNOF introduces the concept of dynamic occupation numbers, as
\begin{equation}
    n_p^\mathrm{d} = n_p \cdot \text{e}^{-\left(\frac{h_g}{h_\mathrm{c}}\right)^2}, \, p\in\Omega_{g}
\end{equation}
with the hole given by $h_g = 1-n_g$ and $h_\mathrm{c} = 0.02\sqrt{2}$. The interpair opposite spin contribution is then given by
\begin{multline}
    \mathrm{D}_{pqrt}^{\alpha\beta\alpha\beta} = \frac{1}{2} \bigg(n_p n_q \delta_{pr} \delta_{qt} \delta_{p\Omega_f} \delta_{q\Omega_g} \\
    - \left[ \Pi^\mathrm{s}_{pr} + \Pi^\mathrm{d}_{pr}  \right]  \delta_{pq} \delta_{rt} \delta_{p\Omega_f} \delta_{r\Omega_g} \bigg) \>\>,
    \label{eq:gnof}    
\end{multline}
with
\begin{multline}
    \Pi^\mathrm{s}_{pr} = \Phi_p \Phi_r ( \delta_{p\Omega^\mathrm{b}}\delta_{r\Omega^\mathrm{a}} + \delta_{p\Omega^\mathrm{a}}\delta_{r\Omega^\mathrm{b}} + \delta_{p\Omega^\mathrm{a}}\delta_{r\Omega^\mathrm{a}} ) \>\>,
\end{multline}
\begin{multline}
    \Pi^\mathrm{d}_{pr} = \left( \sqrt{n_p^\mathrm{d}n_r^\mathrm{d}} - n_p^\mathrm{d} n_r^\mathrm{d}\right) \left( \delta_{p\Omega^\mathrm{b}}\delta_{r\Omega^\mathrm{a}} + \delta_{p\Omega^\mathrm{a}}\delta_{r\Omega^\mathrm{b}} \right)\\ - \left( \sqrt{n_p^\mathrm{d} n_r^\mathrm{d}} + n_p^\mathrm{d}n_r^\mathrm{d} \right) \delta_{p\Omega^\mathrm{a}}  \delta_{r\Omega^\mathrm{a}} \>\>,
\end{multline}
where $\Omega^\mathrm{b}$ denotes the subspace composed of orbitals below the level $N/2$ ($p \le N/2 $), while $\Omega^\mathrm{a}$ denotes the subspace composed of orbitals above the level $N/2$ ($p > N/2 $). Observe that interactions between orbitals belonging to $\Omega^\mathrm{b}$ are not considered in the $\boldsymbol{\Pi}$ matrices of GNOF. 

The matrix $\boldsymbol{\Pi}^\mathrm{d}$ accounts for dynamic correlation between subspaces in accordance with Pulay's criterion, which establishes an occupancy deviation of approximately 0.01 with respect to ``1'' or ``0'' for a natural orbital to contribute to the dynamic correlation, while larger deviations contribute to nondynamic correlation. $\boldsymbol{\Pi}^\mathrm{s}$ from PNOF7 functional form is conserved.

\section{\label{sec:Computational}Computational Details}

The NOF calculations have been carried out using an extended pairing approach, that is, we correlate all electrons into all available orbitals for a given basis set, which today is not possible for large systems with current wavefunction-based methods. The number of weakly double occupied orbitals coupled to each strongly double occupied orbital is indicated as $N_\text{cwo}$ below the plot of each system. CI and TD-DFT calculations were carried out for comparison using the Psi4\cite{Parrish2017-vx,Smith2020-yt} software. All calculations have been performed using a def2-TZVPD basis set,\cite{weigend2005a,rappoport2010a} except for the \ce{N2} calculation that was performed using a cc-pVDZ basis set.\cite{Dunning1989-mo}

The equations of PNOF coupled to ERPA0, ERPA1, and ERPA2 have been implemented in the DoNOF \cite{Piris2021-xo} and in PyNOF software.\cite{PyNOF} It is important to notice that several techniques have been developed to avoid explicit storage of $\bf{A}$ and $\bf{B}$ matrices, as well as the full diagonalization of large matrices.\cite{Olsen1988-kh,Alessandro2023-dk} In particular, the algorithm of Stratmann, Scuseria and Frisch,\cite{Stratmann1998-ox} that take advantage of the fact that the excitation energies appear in pairs, may be applicable to ERPA0 and ERPA1, although with some modifications, as the vectors $\bf{X}+\bf{Y}$ and $\bf{X}-\bf{Y}$ are not orthonormal as in TD-SCF, but instead the orthonormality is hold by the vectors $\bf{X}-\bf{Y}$ and $\bf{\Delta N} (\bf{X}+\bf{Y})$. This can be seen for ERPA0 by rewriting the reduced equations as
\begin{multline}    
    (\bf{A}-\bf{B}) (\bf{\Delta N})^{-1} (\bf{A}+\bf{B}) (\bf{\Delta N})^{-1} \bf{R}\\ = \omega^2 \bf{R} \>\>,
\end{multline}
\begin{multline}    
    (\bf{\Delta N})^{-1} (\bf{A}+\bf{B}) (\bf{\Delta N})^{-1} (\bf{A}-\bf{B}) \bf{L}\\ = \omega^2 \bf{L} \>\>,
\end{multline}
where $\bf{R} = \bf{\Delta N} (\bf{X}+\bf{Y})$ and $\bf{L} = (\bf{X}-\bf{Y})$ are the right and left eigenvectors of the $(\bf{A}-\bf{B}) (\bf{\Delta N})^{-1} (\bf{A}+\bf{B}) (\bf{\Delta N})^{-1}$ matrix. Using this approach would allow to iteratively compute a selected number of excitation by diagonalizing small matrices. A similar approach is applicable to ERPA1. However, some details of the algorithm must be explored, for example, the possibility of $\Delta \bf{N}$ being not invertible, as well as the symmetry of the matrices involved in ERPA1. Furthermore, this approach may not be applicable to ERPA2, as the paired structure of the eigenvalues is lost for this case. For the purpose of this work, we are solving ERPA0 by Eq.~(\ref{eq:reduced-ERPA0-1}), ERPA1 by Eq.~(\ref{eq:reduced-ERPA1-1}) and ERPA2 by Eq.~(\ref{eq:ERPA2}) by performing full diagonalization of the involved matrices.

\section{\label{sec:Results}Results and discussion}

In this section, we present the ground and excited state PECs of model systems, namely \ce{H2}, \ce{HeH+}, \ce{LiH} and \ce{Li2}, computed with PNOFs coupled to ERPA0, ERPA1 and ERPA2. These molecules are of interest due to their low number of electrons that allow the results to be compared directly with the values of the FCI method. In addition, we also present the PEC of \ce{N2}, a larger system which involves breaking a triple bond. 

\subsection{ERPA0 vs ERPA1 vs ERPA2: \ce{H2}, \ce{HeH+} and \ce{LiH}}

The cases of \ce{H2} and \ce{HeH+} are remarkable, since only intrapair (and no interpair) contributions to the electron correlation are required. In these cases, PNOF5, PNOF7, and GNOF converge to the same functional form. We start with the simplest of these molecules, the homonuclear diatomic \ce{H2}. The energies of the ground and excited states are presented in Fig.~\ref{fig:h2-pnof}, with the PNOF results as circle marks and the FCI reference values as solid lines. In particular, Fig.~\ref{fig:h2-pnof-erpa0} and Fig.~\ref{fig:h2-pnof-erpa2} present the results of PNOF-ERPA0 and PNOF-ERPA2, respectively.

\begin{figure}[htb]
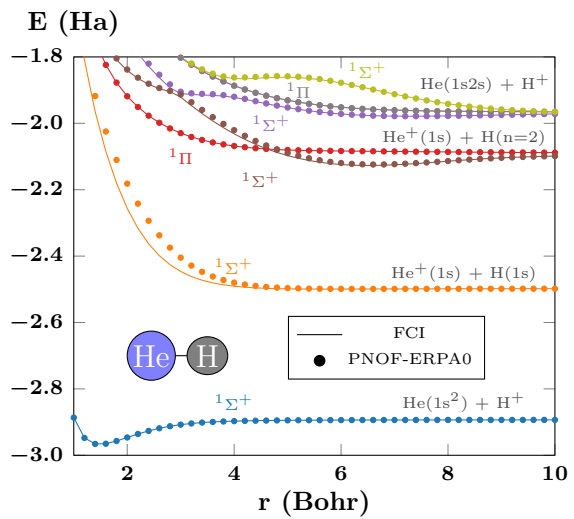

    \centering
    \include{HeH+-ERPA0}
    \vspace{-1.2cm}
    \caption{PECs of the first states of \ce{HeH+} computed using PNOF-ERPA0 and FCI. There are $N_\text{cwo} = 17$   orbitals paired to each strongly double occupied orbital. The first curve corresponds to the ground state.}
    \label{fig:heh+-pnof-erpa0}        
\end{figure}

\begin{figure}[htb]
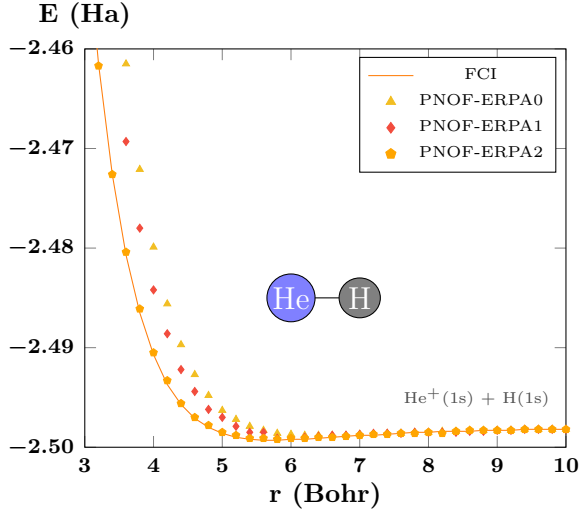

    \centering
    \include{HeH+-ERPAs}
    \vspace{-1.2cm}
    \caption{First excited state of \ce{HeH+}. There are $N_\text{cwo} = 17$  orbitals paired to each strongly double occupied orbital.}
    \label{fig:heh+-pnof-erpas}        
\end{figure}

From Fig.~\ref{fig:h2-pnof-erpa0}, it can be seen that the ground state ($^1 \Sigma_g^+$, blue) and several excited states ($^1 \Sigma_u^+$ orange, $^1 \Pi_u$ red, $^1 \Sigma_u^+$ pink, $^1 \Sigma_u^+$ golden) PECs are in good agreement with FCI. On the other hand, excited states such as those of $^1 \Sigma_g^+$ gray and purple marks agree well with FCI in some but not all the domain, this is caused by a lost state that makes it impossible to capture the avoided crossing between the gray and purple curves around 2.6 Bohr. These deviations can be tracked to violation of the consistency condition. In this regard, PNOF-ERPA1, which includes single diagonal excitations, provides almost the same results as PNOF-ERPA0, and the problem can only be solved by including diagonal double excitations.\cite{Chatterjee2012-qn} The results of PNOF-ERPA2 presented in Fig.~\ref{fig:h2-pnof-erpa2} show that it can recover the lost state; consequently, the avoided crossing and the shape of the gray and purple curves are well described. It should be noted that, even though PNOF-ERPA0 loses the avoided crossing, it is still able to describe the intersection between the pink and gray curves at 3.2 Bohr.

\begin{figure}[htb]
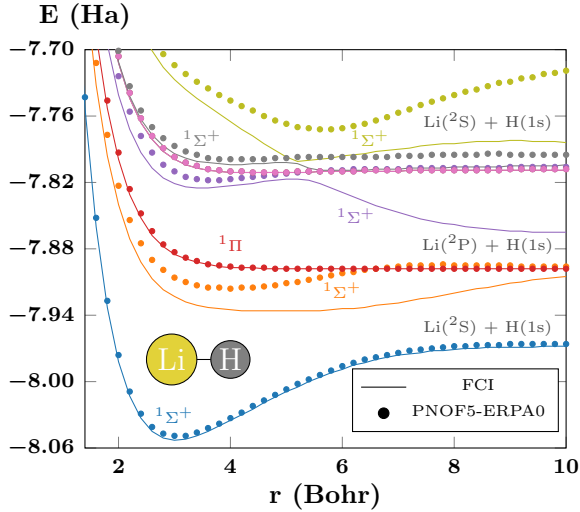

    \centering
    \include{LiH-PNOF5-ERPA0}
    \vspace{-1.2cm}
    \caption{PECs of the excited states of \ce{LiH} computed using PNOF5-ERPA0 and FCI. There are $N_\text{cwo} = 12$  orbitals paired to each strongly double occupied orbital. The first curve corresponds to the ground state.}
    \label{fig:lih-pnof5-erpa0}        
\end{figure}

\begin{figure*}[htb]
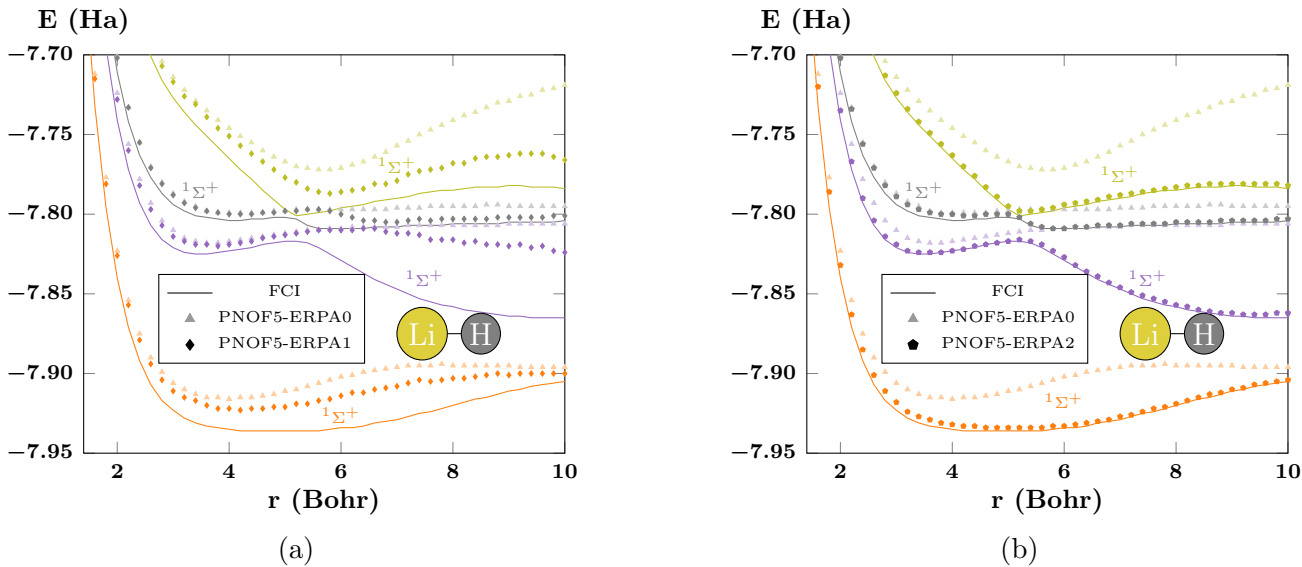

    \centering
    \begin{subfigure}[b]{0.45\textwidth}
    \include{LiH-PNOF5-ERPA}
    \vspace{-1.0cm}
    \caption{}
    \label{fig:lih-pnof5-erpa}        
    \end{subfigure}
    \hfill
    \centering
    \begin{subfigure}[b]{0.45\textwidth}
    \include{LiH-PNOF5-ERPA2}
    \vspace{-1.0cm}
    \caption{}
    \label{fig:lih-pnof5-erpa2}        
    \end{subfigure}
    \hfill
    \caption{PECs of selected states of \ce{LiH} computed using a) PNOF5-ERPA1 and b) PNOF5-ERPA2. FCI results are presented as solid lines, and PNOF5-ERPA0 values are shown as attenuated triangles on the background for comparison. There are $N_\text{cwo} = 12$ orbitals paired to each strongly double occupied orbital. The ground state is not shown.}
\end{figure*}
\begin{figure*}[htb]
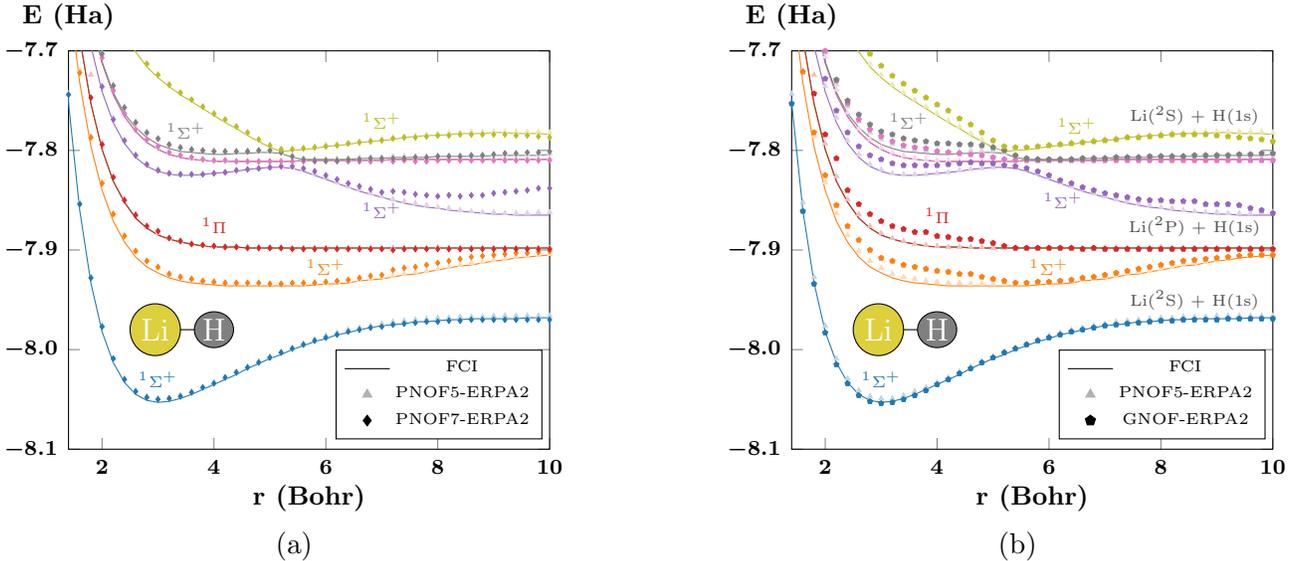

    \centering
    \begin{subfigure}[t]{0.45\textwidth}
    \include{LiH-PNOF7-ERPA2}
    \vspace{-1.0cm}
    \caption{}
    \label{fig:lih-pnof7-erpa2}
    \end{subfigure}
    \hfill
    \centering
    \begin{subfigure}[t]{0.45\textwidth}
    \include{LiH-GNOF-ERPA2}
    \vspace{-1.0cm}
    \caption{}
    \label{fig:lih-gnof-erpa2}
    \end{subfigure}
    \hfill
    \caption{PECs of the first states of \ce{LiH} computed using a) PNOF7-ERPA2 and b) GNOF-ERPA2. There are $N_\text{cwo} = 12$ orbitals paired to each strongly double occupied orbital. The ground state is shown in blue.}
\end{figure*}
On the other hand, Fig.~\ref{fig:heh+-pnof-erpa0} shows the PECs of \ce{HeH+}, a diatomic heteronuclear charged system, computed with PNOF-ERPA0. It can be seen that most of the results accurately reproduce the FCI values, including the avoided crossing between the $^1 \Sigma^+$ brown and purple curves at 3.0 Bohr, and the crossing between the $^1 \Sigma^+$ brown and $^1 \Pi$ red curves at 4.6 Bohr. In this case, no state has been lost, although the $^1 \Sigma^+$ orange curve that corresponds to the first excitation exhibits some deviations at a distance separation below 5 Bohr. This can be improved, as can be seen in Fig.~\ref{fig:heh+-pnof-erpas}, where going from ERPA0 (triangles) to ERPA1 (diamonds) provides better results, and going to ERPA2 (pentagons) makes the values accurate. It is worth noting that no significant differences are observed beyond 6 Bohr of bond length.

The lithium hydride, with two electron pairs, represents a more correlated system, being the first system in this work to present interpair electron correlation. The energies of the first states of \ce{LiH} calculated with PNOF5-ERPA0 are presented in Fig.~\ref{fig:lih-pnof5-erpa0}, where it can be seen that the method is capable of capture the general profile of the PECs. However, there are some details worth discussing, especially since this system has been used as a model due to its well-known avoided crossings.\cite{Grofe2017-ke} 

The $^1\Sigma^{+}$ orange curve tends to increase in energy too soon as the molecule is dissociated, and the main deviation occurs around 7 Bohr, where a strongly avoided crossing between the $^1\Sigma^{+}$ orange and blue curves occurs. Similarly, GNOF-ERPA0 does not describe well the avoided crossing between the $^1\Sigma^{+}$ orange and purple curves at 10 Bohr, and the avoided crossing between the $^1\Sigma^{+}$ purple and gray curves at around 5.4 Bohr. The curves involved in these avoided crossings are affected in their energy predictions. Moving from ERPA0, to ERPA1 and ERPA2 does not significantly affect the points that are already accurate in Fig.~\ref{fig:lih-pnof5-erpa0}, that is, those of the blue, red, and pink curves, and in any case improves slightly their accuracy; then, in the following, we will remove these curves and focus on the other PECs for clarity. 

Fig.~\ref{fig:lih-pnof5-erpa} presents the improved curves achieved by PNOF5-ERPA1 (diamonds), and the values of PNOF5-ERPA0 (triangles) are shown attenuated on the background as a reference. It can be seen that ERPA1 improves the ERPA0 results by allowing the $^1\Sigma^{+}$ gray curve to become closer to the FCI reference. The $^1\Sigma^{+}$ orange, golden, and gray curves are qualitatively improved by taking the appropriate shape. Although PNOF5-ERPA1 is not completely accurate, it shows that, in this case, the single diagonal excitations may become significant to go beyond the PNOF5-ERPA0 approximation. Furthermore, the PNOF5-ERPA2 approach is capable of recovering all avoided crossings in the studied region, as shown in Fig.~\ref{fig:lih-pnof5-erpa2}, with marks that are very close to the FCI results. It is clear that going from ERPA0 to ERPA1 and ERPA2 improves the results as the marks become closer to the lines of the FCI reference.

\subsection{PNOF5 vs PNOF7 vs GNOF: \ce{LiH} and \ce{Li2}}

Since \ce{LiH} has interpair electron correlation, PNOF5, PNOF7, and GNOF provide different results, as previously reported for the ground state,\cite{Lew-Yee2023-vf} with PNOF5 presenting the highest energies, GNOF being close to FCI, and PNOF7 remaining at intermediate energies; although the energy differences are small. Regarding the excited states, as for this system the dynamic correlation is dominant around the binding region, the PNOF7-ERPA2 PECs show no significant differences with those of PNOF5-ERPA2, but present some deviations in the dissociation region beyond 7.0 Bohr, as can be seen in Fig.~\ref{fig:lih-pnof7-erpa2}. On the other hand, the GNOF-ERPA2 picture also resembles that of PNOF5-ERPA2 but with some small deviations near the binding region, as can be seen in Fig.~\ref{fig:lih-gnof-erpa2}. This is most evident for the $^1 \Sigma^+$ orange and $^1\Pi$ red curves around 4 Bohr of interatomic distances. The pointed discrepancies could be related to the fact that PNOF5 is strictly N-representable, while PNOF7 and GNOF only satisfy some necessary N-representability conditions. Due to the low number of electrons in \ce{LiH}, the violations of the N-representability appear to have a more significant contribution over the consideration of the interpair electron correlation. However, this relation changes as the number of electrons increases, as will be seen below.

The effect of the interpair electron correlation becomes more evident in \ce{Li2}, with three electron pairs. As the behavior of ERPA0, ERPA1 and ERPA2 has been established, here we use ERPA2 directly and look for the difference between PNOFs. The PECs of \ce{Li2} computed with PNOF5-ERPA2 are presented in Fig.~\ref{fig:Li2-pnof5-erpa2}, where it can be seen that PNOF5-ERPA2 achieves qualitatively good results. However, although the curves have been recovered, further inspection shows deviations in almost all cases relative to the FCI lines. Going beyond the independent-pair model becomes important, as can be seen in Fig.~\ref{fig:Li2-gnof-erpa2} where the excited state energies computed with GNOF-ERPA2 are presented. In this case, all the marks become closer to the FCI curves, especially those corresponding to the blue, purple, and golden $^1 \Sigma_\text{g}^+$ curves, as well as the gray $^1 \Sigma_\text{u}^+$ PECs, thus providing more accurate values due to the better treatment of the electron correlation.

\begin{figure*}[t]
    \centering
    \begin{subfigure}[t]{0.45\textwidth}
    \include{Li2-PNOF5-ERPA2}
    \vspace{-1.2cm}
    \caption{}
    \label{fig:Li2-pnof5-erpa2}        
    \end{subfigure}
    \centering
    \begin{subfigure}[t]{0.45\textwidth}
    \include{Li2-GNOF-ERPA2}
    \vspace{-1.2cm}
    \caption{}
    \label{fig:Li2-gnof-erpa2}        
    \end{subfigure}
    \caption{PECs of the first states of \ce{Li2} computed using a) PNOF5-ERPA2 and b) GNOF-ERPA2. There are $N_\text{cwo} = 10$  orbitals paired to each strongly double occupied orbital. The first curve corresponds to the ground state.}
\end{figure*}

\begin{figure}[t]
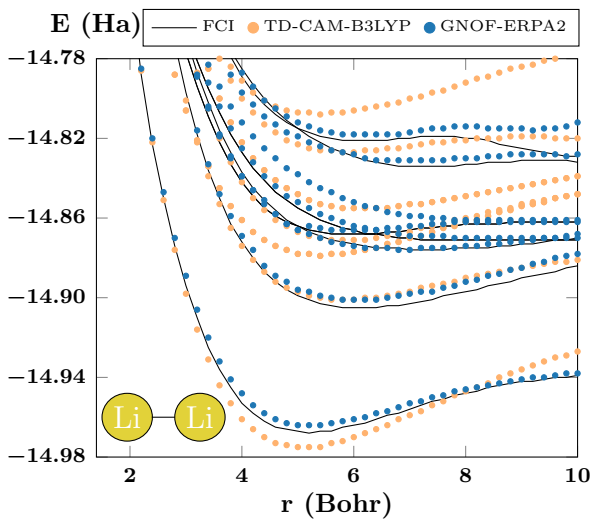

    \centering
    \include{Li2-References}
    \vspace{-1.2cm}
    \caption{PECs of the first states of \ce{Li2} computed using FCI, TD-CAM-B3LYP, and GNOF-ERPA2. There are $N_\text{cwo} = 10$  orbitals paired to each strongly double occupied orbital.}
    \label{fig:Li2-references}        
\end{figure}

Finally, Fig.~\ref{fig:Li2-references} presents a comparison of the results of GNOF-ERPA2 (blue circles) with those of TD-CAM-B3LYP (orange circles) and FCI (black lines). It can be seen that TD-CAM-B3LYP recovers a certain resemblance to the profile of the FCI curves, but there is a significant quantitative difference in favor of GNOF-ERPA2. Moreover, TD-CAM-B3LYP PECs show an overestimation of the electron correlation in the bonding region, but the PECs cross the lines of FCI as the correlation is underestimated in the dissociation process. This can be improved but not completely corrected by an unrestricted TD-CAM-B3LYP calculation, as shown in the SI, at the price of spin contamination. A great advantage of GNOF is that its PECs tend to be parallel to the FCI PECs as a consequence of the balanced treatment of electron correlation, this without introducing spin contamination.

\begin{figure}[htb]
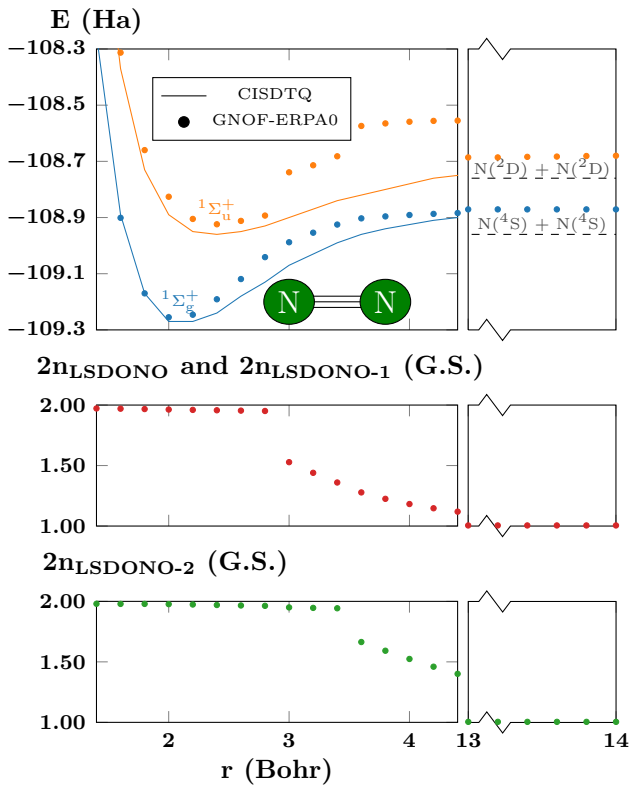

    \centering
    \include{N2}
    \vspace{-1.2cm}
    \caption{PECs of \ce{N2} computed using GNOF-ERPA0 and CISDTQ. There are $N_\text{cwo} = 3$ orbitals paired to each strongly double occupied natural orbital. The top panel present the energy of the ground and the first-excited state. The dotted lines corresponds to the FCI energy of two N($^4$S) atoms and two N($^2$D) atoms. The middle panel corresponds to the occupation numbers of the LSDONO and LSDONO-1, and the bottom panel correspond to the occupation numbers of the LSDONO-2 for the ground state. }
    \label{fig:n2-pec}        
\end{figure}

\subsection{Multiple Bonds: \ce{N2}}

The molecular nitrogen provides a challenging system, as a triple bond is involved, allowing to test the capabilities and limitations of the current PNOF-ERPA implementation. The PECs of the ground state $^1\Sigma_\text{g}^+$ and of the first excited state $^1\Sigma_\text{u}^+$ were calculated using a cc-pVDZ basis set and are presented in the top panel of Fig.~\ref{fig:n2-pec}, with the solid lines corresponding to a CISDTQ calculation that is very close to the reported values of FCI in the bonding region,\cite{Larsen2000-mi} and the circle marks corresponding to the GNOF-ERPA0 results. The FCI values of the ground and the first excited state at the dissociation limit are indicated in dashed lines.

In order to analyze the results it is convenient to divide the dissociation in three zones, the first corresponding to a separation distance below 2.8 Bohr and containing the bonding region, characterized by occupation numbers close to ``two'' for the strongly double occupied natural orbitals. The second region corresponds to the interval between 2.8 Bohr and 3.6 Bohr, and is characterized by the lowest strongly double occupied natural orbitals (LSDONO) becoming fractional occupied as can be seen in the red curve at the middle panel. A similar behavior is obtained for the LSDONO-1, which together with LSDONO represents the bond breaking process of the two $\pi$ orbitals. The third region appears for distances beyond 3.6 Bohr, and is characterized by the occupation numbers of the LSDONO-2 becoming fractional, as can be seen in the green curve at the top panel of Fig.~\ref{fig:n2-pec}. This time, the process corresponds to the breaking of the $\sigma$ orbital. Finally, as the separation distance of the nitrogen atoms increases, the occupation numbers of LSDONO, LSDONO-1 and LSDONO-2 move to values close to ``one'', which together with the coupled weakly occupied natural orbitals represents the complete dissociation to two N($^4$S) atoms.

Regarding the first region, GNOF achieves a remarkable success by providing by itself energies that are very close to the CI results for both the ground state and the first excited state. For the second and third regions, the ground state predicted by GNOF remains close to the CI results, although the change of the occupation numbers at 2.8 Bohr for the LSDONO and LSDONO-1, and at 3.6 Bohr for the LSDONO-2 is not smooth. This behavior of GNOF is already known when moving from electron correlation regime,\cite{Mitxelena2022-fs} and is reflected in the first excited state that presents non-smooth transitions exactly in these values of the separation distance.

Finally, it is important to mention that achieving the correct excitation energies in the second and particularly in the third region is difficult due to the fact that there are occupation numbers with the same value, for example those of the LSDONO and the LSDONO-1. This is particularly significant at the dissociation limit, where there are six occupation numbers with almost the same value of ``one''; therefore, the $\bf{\Delta N}$ matrix presents several zeros and becomes non-invertible. On the right side of the plot, we present selected points of the first excited state. GNOF provides an excitation energy of 0.19 Ha in good agreement with the value of 0.20 Ha provided by FCI. However, we still want to highlight that the current algorithm becomes unestable in this scenario. We attribute these difficulties not to inaccuracies in the GNOF-ERPA approach, but to the fact that the $\bm{\Delta N}$ matrix may be ill-conditioned and that several algebraic techniques should be explored for these cases.

\section{\label{sec:Conclusions}Conclusions}

This work validates the coupling of PNOF functionals with the ERPA equations as a very promising approach for studying excited states. As expected, going from ERPA0 to ERPA1 and ERPA2 improves the results. It is important to note that ERPA0 have shown inaccuracies regarding avoided crossings in the studied systems, and although ERPA1 improve the results, ERPA2 have been required to describe them correctly. Despite this fact, ERPA0 has been able to correctly describe crossings between curves and provides a general depiction of the excited states.

Regarding the functionals tested, PNOF5 seems to be enough for small molecules, but as the size of the systems increases, the interpair electron correlation becomes important and PNOF7 and GNOF provide better results. The PNOF-ERPA approach becomes promising in the context of the other methods used for excited states, as PNOF provides values comparable to those of high-level CI. We must recall that the cost of a ground state PNOF calculation is of the fourth order with the number of orbitals for the ground state, while the cost of calculating the CI wavefunction depends on the number of determinants with exponential growth. In fact, the scaling of the ground-state PNOF calculation is comparable with that of hybrid density functional approximations, with the advantage of the PNOF of being able to deal simultaneously with charge delocalization and static correlation. Once the ground-state PNOF result has been achieved, the scaling of the excited-state calculation becomes of the sixth order, comparable to that of standard TD-DFT, but with substantially better results, as has been shown in this work.

The capabilities of PNOF calculations have now been extended to all chemical problems that involve excited states; for example, in the future the study of photochemical processes may benefit from a balanced inclusion of static and dynamic correlation. It is expected that the accuracy of the excited states will be greatly benefited by the development of better functionals that surpass the currently good performance of the GNOF. Finally, as the potential of the PNOF-ERPA approach has been stablished, it is desirable to develop the implementation that avoid the diagonalization of the full matrix, as well as taken care of challenging cases with degeneracy on the values of the occupation numbers.

\begin{acknowledgement}

J. F. H. Lew-Yee with CVU Grant No. 867718 acknowledges ``Consejo Nacional de Ciencia y Tecnología (CONACyT)'' and ``Universidad Nacional Autónoma de México (UNAM)'' for the Ph.D. scholarship. J. M. del Campo acknowledges funding with project Grant No. IN201822 from PAPIIT, and computing resources from ``Laboratorio Nacional de Cómputo de Alto Desempeño (LANCAD)'' with project Grant No. LANCAD-UNAMDGTIC-270. M. Piris acknowledges funding from MCIN/AEI/10.13039/501100011033 (Ref.: PID2021-126714NB-I00) and the Eusko Jaurlaritza (Ref.: IT1584-22). 

\end{acknowledgement}


\providecommand{\latin}[1]{#1}
\makeatletter
\providecommand{\doi}
  {\begingroup\let\do\@makeother\dospecials
  \catcode`\{=1 \catcode`\}=2 \doi@aux}
\providecommand{\doi@aux}[1]{\endgroup\texttt{#1}}
\makeatother
\providecommand*\mcitethebibliography{\thebibliography}
\csname @ifundefined\endcsname{endmcitethebibliography}  {\let\endmcitethebibliography\endthebibliography}{}

\end{document}

%% file: pairing_scheme.tex
\begin{tikzpicture}

\node[color1] at (2.1,0.3) {$\Omega_1$};
\draw[<->,color1, very thick] (2.0,5.25) -- (3.1,5.25) -- (3.1,0.0) -- (2.0,0.0);
\draw[color1, thick,dashed] (-0.2,4.7) rectangle (1.7,6.0);
\draw[<-, very thick] (1.0,5.6) -- (1.0,5.75);
\draw[->, very thick] (0.5,5.6) -- (0.5,5.75);
\draw[color1, very thick] (0.0,5.6) -- (1.5,5.6);
\draw[<-, very thick] (1.0,4.9) -- (1.0,5.05);
\draw[->, very thick] (0.5,4.9) -- (0.5,5.05);
\draw[color1, very thick] (0.0,4.9) -- (1.5,4.9);

\node[color2] at (2.1,1.0) {$\Omega_2$};
\draw[<->,color2, very thick] (2.0,3.85) -- (2.8,3.85) -- (2.8,0.7) -- (2.0,0.7);
\draw[color2, thick, dashed] (-0.2,3.3) rectangle (1.7,4.6);
\draw[<-, very thick] (1.0,4.2) -- (1.0,4.40);
\draw[->, very thick] (0.5,4.2) -- (0.5,4.40);
\draw[color2, very thick] (0.0,4.2) -- (1.5,4.2);
\draw[<-, very thick] (1.0,3.5) -- (1.0,3.70);
\draw[->, very thick] (0.5,3.5) -- (0.5,3.70);
\draw[color2, very thick] (0.0,3.5) -- (1.5,3.5);

\node[color3] at (2.1,1.7) {$\Omega_3$};
\draw[<->,color3, very thick] (2.0,2.45) -- (2.5,2.45) -- (2.5,1.4) -- (2.0,1.4);
\draw[color3, thick, dashed] (-0.2,1.9) rectangle (1.7,3.2);
\draw[<-, very thick] (1.0,2.8) -- (1.0,3.05);
\draw[->, very thick] (0.5,2.8) -- (0.5,3.05);
\draw[color3, very thick] (0.0,2.8) -- (1.5,2.8);
\draw[<-, very thick] (1.0,2.1) -- (1.0,2.35);
\draw[->, very thick] (0.5,2.1) -- (0.5,2.35);
\draw[color3, very thick] (0.0,2.1) -- (1.5,2.1);

\draw[<-, very thick] (1.0,1.4) -- (1.0,1.7);
\draw[->, very thick] (0.5,1.4) -- (0.5,1.7);
\draw[color3, very thick] (0.0,1.4) -- (1.5,1.4);

\draw[<-, very thick] (1.0,0.7) -- (1.0,1.1);
\draw[->, very thick] (0.5,0.7) -- (0.5,1.1);
\draw[color2, very thick] (0.0,0.7) -- (1.5,0.7);

\draw[<-, very thick] (1.0,0.0) -- (1.0,0.5);
\draw[->, very thick] (0.5,0.0) -- (0.5,0.5);
\draw[color1, very thick] (0.0,0.0) -- (1.5,0.0);

\node at (-2.5,4.15) {Weakly Double};
\node at (-2.5,3.55) {Occupied Orbitals};

\node at (-2.5,1.3) {Strongly Double};
\node at (-2.5,0.7) {Occupied Orbitals};

\end{tikzpicture}

%% file: H2-ERPA0.tex
\begin{tikzpicture}
\begin{axis}[
    width=\textwidth,
    enlargelimits=false,
    cycle list name = matplotlib,
    xmin = 1.0,
    xmax = 8.0,    
    ymax = -0.4,
    ymin = -1.2,
    xlabel = r (Bohr),
    ylabel = E (Ha),
    y tick label style={
        /pgf/number format/.cd,
        fixed,
        fixed zerofill,
        precision=1,
        /tikz/.cd
    },
    every axis y label/.style={
    at={(ticklabel* cs:1.05)},
    anchor=south,
    },
    legend style={/tikz/every even column/.append style={column sep=0.5cm},at={(0.65,0.52)},anchor=north,font=\tiny},
    ylabel style={align=center, inner sep=0pt, font=\footnotesize\bfseries\boldmath},
    xlabel style={align=center, inner sep=0pt, font=\footnotesize\bfseries\boldmath},
    x tick label style={font=\scriptsize\bfseries\boldmath},
    y tick label style={font=\scriptsize\bfseries\boldmath},
]

\addplot+[steelblue31119180,forget plot] table[x=r, y=E0] {h2-fci.dat};
\addplot+[darkorange25512714,forget plot] table[x=r, y=E1] {h2-fci.dat};
\addplot+[crimson2143940,forget plot] table[x=r, y=E3] {h2-fci.dat};
\addplot+[mediumpurple148103189,forget plot] table[x=r, y=E4] {h2-fci.dat};
\addplot+[orchid227119194,forget plot] table[x=r, y=E6] {h2-fci.dat};
\addplot+[gray127,forget plot] table[x=r, y=E7] {h2-fci.dat};
\addplot+[goldenrod18818934,forget plot] table[x=r, y=E8] {h2-fci.dat};

\addplot[] coordinates {(1,1)(2,2)};
\addlegendentry{FCI}

\addplot+[steelblue31119180,only marks,mark size=1pt,forget plot] table[x=r, y=E0] {h2-pnof-erpa0.dat};
\addplot+[darkorange25512714,only marks,mark size=1pt,forget plot] table[x=r, y=E1] {h2-pnof-erpa0.dat};
\addplot+[crimson2143940,only marks,mark size=1pt,forget plot] table[x=r, y=E3] {h2-pnof-erpa0.dat};
\addplot+[gray127,only marks,mark size=1pt,forget plot] table[x=r, y=E4] {h2-pnof-erpa0.dat};
\addplot+[orchid227119194,only marks,mark size=1pt,forget plot] table[x=r, y=E6] {h2-pnof-erpa0.dat};
\addplot+[goldenrod18818934,only marks,mark size=1pt,forget plot] table[x=r, y=E7] {h2-pnof-erpa0.dat};

\addplot[only marks,mark size=2pt] coordinates {(1,1)(2,2)};
\addlegendentry{PNOF-ERPA0}

\node[steelblue31119180] at (1.49,-1.11) {\tiny $^1\Sigma^{+}_\text{g}$};
\node[darkorange25512714] at (2.4,-0.705) {\tiny $^1\Sigma^{+}_\text{u}$};
\node[crimson2143940] at (1.9,-0.63) {\tiny $^1\Pi_\text{u}$};
\node[mediumpurple148103189] at (4.0,-0.66) {\tiny $^1\Sigma^{+}_\text{g}$};
\node[orchid227119194] at (6,-0.63) {\tiny $^1\Sigma^{+}_\text{u}$};
\node[gray127] at (2,-0.45) {\tiny $^1\Sigma^{+}_\text{g}$};
\node[goldenrod18818934] at (3,-0.49) {\tiny $^1\Sigma^{+}_\text{u}$};

\node[black!70] at (6.8,-0.52) {\tiny H(n$=$2) + H(1s)};
\node[black!70] at (6.8,-1.04) {\tiny H(1s) + H(1s)};

\draw[fill=gray!100,text=white] (2.0,-0.9) ellipse (0.35 and 0.05) node {H};
\draw (2.35,-0.9) -- (2.65,-0.9);
\draw[fill=gray!100,text=white] (3.0,-0.9) ellipse (0.35 and 0.05) node {H};

\end{axis}
\end{tikzpicture}

%% file: H2-ERPA2.tex
\begin{tikzpicture}
\begin{axis}[
    width=\textwidth,
    enlargelimits=false,
    cycle list name = matplotlib,
    xmin = 1.0,
    xmax = 8.0,
    ymax = -0.4,
    ymin = -1.2,
    xlabel = r (Bohr),
    ylabel = E (Ha),
    y tick label style={
        /pgf/number format/.cd,
        fixed,
        fixed zerofill,
        precision=1,
        /tikz/.cd
    },
    every axis y label/.style={
    at={(ticklabel* cs:1.05)},
    anchor=south,
    },
    legend style={/tikz/every even column/.append style={column sep=0.5cm},at={(0.65,0.52)},anchor=north,font=\tiny},
    ylabel style={align=center, inner sep=0pt, font=\footnotesize\bfseries\boldmath},
    xlabel style={align=center, inner sep=0pt, font=\footnotesize\bfseries\boldmath},
    x tick label style={font=\scriptsize\bfseries\boldmath},
    y tick label style={font=\scriptsize\bfseries\boldmath},
]

\addplot+[steelblue31119180,forget plot] table[x=r, y=E0] {h2-fci.dat};
\addplot+[darkorange25512714,forget plot] table[x=r, y=E1] {h2-fci.dat};
\addplot+[crimson2143940,forget plot] table[x=r, y=E3] {h2-fci.dat};
\addplot+[mediumpurple148103189,forget plot] table[x=r, y=E4] {h2-fci.dat};
\addplot+[orchid227119194,forget plot] table[x=r, y=E6] {h2-fci.dat};
\addplot+[gray127,forget plot] table[x=r, y=E7] {h2-fci.dat};
\addplot+[goldenrod18818934,forget plot] table[x=r, y=E8] {h2-fci.dat};

\addplot[] coordinates {(1,1)(2,2)};
\addlegendentry{FCI}

\addplot+[steelblue31119180,only marks,mark size=1pt,forget plot] table[x=r, y=E0] {h2-pnof-erpa2.dat};
\addplot+[darkorange25512714,only marks,mark size=1pt,forget plot] table[x=r, y=E1] {h2-pnof-erpa2.dat};
\addplot+[crimson2143940,only marks,mark size=1pt,forget plot] table[x=r, y=E3] {h2-pnof-erpa2.dat};
\addplot+[mediumpurple148103189,only marks,mark size=1pt,forget plot] table[x=r, y=E4] {h2-pnof-erpa2.dat};
\addplot+[orchid227119194,only marks,mark size=1pt,forget plot] table[x=r, y=E6] {h2-pnof-erpa2.dat};
\addplot+[gray127,only marks,mark size=1pt,forget plot] table[x=r, y=E7] {h2-pnof-erpa2.dat};
\addplot+[goldenrod18818934,only marks,mark size=1pt,forget plot] table[x=r, y=E8] {h2-pnof-erpa2.dat};

\addplot[only marks,mark size=2pt] coordinates {(1,1)(2,2)};
\addlegendentry{PNOF-ERPA2}

\draw[fill=gray!100,text=white] (2.0,-0.9) ellipse (0.35 and 0.05) node {H};
\draw (2.35,-0.9) -- (2.65,-0.9);
\draw[fill=gray!100,text=white] (3.0,-0.9) ellipse (0.35 and 0.05) node {H};

\node[black!70] at (6.8,-0.52) {\tiny H(n$=$2) + H(1s)};
\node[black!70] at (6.8,-1.04) {\tiny H(1s) + H(1s)};

\node[steelblue31119180] at (1.49,-1.11) {\tiny $^1\Sigma^{+}_\text{g}$};
\node[darkorange25512714] at (2.4,-0.705) {\tiny $^1\Sigma^{+}_\text{u}$};
\node[crimson2143940] at (1.9,-0.63) {\tiny $^1\Pi_\text{u}$};
\node[mediumpurple148103189] at (4.0,-0.66) {\tiny $^1\Sigma^{+}_\text{g}$};
\node[orchid227119194] at (6,-0.63) {\tiny $^1\Sigma^{+}_\text{u}$};
\node[gray127] at (2,-0.45) {\tiny $^1\Sigma^{+}_\text{g}$};
\node[goldenrod18818934] at (3,-0.49) {\tiny $^1\Sigma^{+}_\text{u}$};

\end{axis}
\end{tikzpicture}

%% file: HeH+-ERPA0.tex
\begin{tikzpicture}
\begin{axis}[
    width=0.45\textwidth,
    enlargelimits=false,
    cycle list name = matplotlib,
    xmin = 1.0,
    ymax = -1.8,
    ymin = -3.0,
    xlabel = r (Bohr),
    ylabel = E (Ha),
    y tick label style={
        /pgf/number format/.cd,
        fixed,
        fixed zerofill,
        precision=1,
        /tikz/.cd
    },
    every axis y label/.style={
    at={(ticklabel* cs:1.05)},
    anchor=south,
    },
    legend style={/tikz/every even column/.append style={column sep=0.5cm},at={(0.65,0.35)},anchor=north,font=\tiny},
    ylabel style={align=center, inner sep=0pt, font=\footnotesize\bfseries\boldmath},
    xlabel style={align=center, inner sep=0pt, font=\footnotesize\bfseries\boldmath},
    x tick label style={font=\scriptsize\bfseries\boldmath},
    y tick label style={font=\scriptsize\bfseries\boldmath},
]

\addplot+[steelblue31119180,forget plot] table[x=r, y=E0] {heh+-fci.dat};
\addplot+[darkorange25512714,forget plot] table[x=r, y=E1] {heh+-fci.dat};
\addplot+[crimson2143940,forget plot] table[x=r, y=E3] {heh+-fci.dat};
\addplot+[mediumpurple148103189,forget plot] table[x=r, y=E4] {heh+-fci.dat};
\addplot+[sienna1408675,forget plot] table[x=r, y=E5] {heh+-fci.dat};
\addplot+[orchid227119194,forget plot] table[x=r, y=E6] {heh+-fci.dat};
\addplot+[gray127,forget plot] table[x=r, y=E7] {heh+-fci.dat};
\addplot+[goldenrod18818934,forget plot] table[x=r, y=E8] {heh+-fci.dat};

\addplot[] coordinates {(1,1)(2,2)};
\addlegendentry{FCI}

\addplot+[steelblue31119180,only marks,mark size=1pt,forget plot] table[x=r, y=E0] {heh+-pnof-erpa0.dat};
\addplot+[darkorange25512714,only marks,mark size=1pt,forget plot] table[x=r, y=E1] {heh+-pnof-erpa0.dat};
\addplot+[crimson2143940,only marks,mark size=1pt,forget plot] table[x=r, y=E3] {heh+-pnof-erpa0.dat};
\addplot+[mediumpurple148103189,only marks,mark size=1pt,forget plot] table[x=r, y=E4] {heh+-pnof-erpa0.dat};
\addplot+[sienna1408675,only marks,mark size=1pt,forget plot] table[x=r, y=E5] {heh+-pnof-erpa0.dat};
\addplot+[orchid227119194,only marks,mark size=1pt,forget plot] table[x=r, y=E6] {heh+-pnof-erpa0.dat};
\addplot+[gray127,only marks,mark size=1pt,forget plot] table[x=r, y=E7] {heh+-pnof-erpa0.dat};
\addplot+[goldenrod18818934,only marks,mark size=1pt,forget plot] table[x=r, y=E8] {heh+-pnof-erpa0.dat};

\addplot[only marks,mark size=2pt] coordinates {(1,1)(2,2)};
\addlegendentry{PNOF-ERPA0}

\draw[fill=blue!50,text=white] (2.45,-2.7) ellipse (0.45 and 0.074) node {He};
\draw (2.90,-2.7) -- (3.15,-2.7);
\draw[fill=gray!100,text=white] (3.5,-2.7) ellipse (0.38 and 0.057) node {H};

\node[black!70] at (8.7,-1.88) {\tiny He(1s2s) + H$^+$};
\node[black!70] at (8.3,-2.04) {\tiny \ce{He+}(1s) + H(n=2)};
\node[black!70] at (8.3,-2.45) {\tiny \ce{He+}(1s) + H(1s)};
\node[black!70] at (8.3,-2.84) {\tiny He(1s$^2$) + H$^+$};

\node[steelblue31119180] at (4,-2.84) {\tiny $^1\Sigma^{+}$};
\node[darkorange25512714] at (4.0,-2.43) {\tiny $^1\Sigma^{+}$};
\node[crimson2143940] at (3,-2.1) {\tiny $^1\Pi$};
\node[mediumpurple148103189] at (4.7,-2.0) {\tiny $^1\Sigma^{+}$};
\node[sienna1408675] at (4.5,-2.17) {\tiny $^1\Sigma^{+}$};
\node[gray127] at (5.2,-1.9) {\tiny $^1\Pi$};
\node[goldenrod18818934] at (6.5,-1.84) {\tiny $^1\Sigma^{+}$};

\end{axis}
\end{tikzpicture}

%% file: HeH+-ERPAs.tex
\begin{tikzpicture}
\begin{axis}[
    width=0.45\textwidth,
    enlargelimits=false,
    cycle list name = matplotlib,
    xmin = 3.0,
    ymax = -2.46,
    ymin = -2.50,
    xlabel = r (Bohr),
    ylabel = E (Ha),
    y tick label style={
        /pgf/number format/.cd,
        fixed,
        fixed zerofill,
        precision=2,
        /tikz/.cd
    },
    every axis y label/.style={
    at={(ticklabel* cs:1.05)},
    anchor=south,
    },
    legend style={font=\tiny},
    ytick distance={0.01},
    ylabel style={align=center, inner sep=0pt, font=\footnotesize\bfseries\boldmath},
    xlabel style={align=center, inner sep=0pt, font=\footnotesize\bfseries\boldmath},
    x tick label style={font=\scriptsize\bfseries\boldmath},
    y tick label style={font=\scriptsize\bfseries\boldmath},
]

\addplot+[darkorange25512714] table[x=r, y=E1] {heh+-fci.dat};
\addlegendentry{FCI}

\addplot+[orange1,only marks,mark size=1.5pt,mark=triangle*,forget plot] table[x=r, y=E1] {heh+-pnof-erpa0.dat};
\addplot[orange1,only marks,mark size=2pt,mark=triangle*] coordinates {(1,-2.60)(2,-2.60)};
\addlegendentry{PNOF-ERPA0}

\addplot+[orange2,only marks,mark size=1.5pt,mark=diamond*,forget plot] table[x=r, y=E1] {heh+-pnof-erpa.dat};
\addplot[orange2,only marks,mark size=2pt,mark=diamond*] coordinates {(1,-2.60)(2,-2.60)};
\addlegendentry{PNOF-ERPA1}

\addplot+[orange3,only marks,mark size=1.5pt,mark=pentagon*,forget plot] table[x=r, y=E1] {heh+-pnof-erpa2.dat};
\addplot[orange3,only marks,mark size=2pt,mark=pentagon*] coordinates {(1,-2.60)(2,-2.60)};
\addlegendentry{PNOF-ERPA2}


\node[black!70] at (8.7,-2.495) {\tiny \ce{He+}(1s) + H(1s)};

\draw[fill=blue!50,text=white] (6.0,-2.485) ellipse (0.35 and 0.0024) node {He};
\draw (6.35,-2.485) -- (6.70,-2.485);
\draw[fill=gray!100,text=white] (7.0,-2.485) ellipse (0.3 and 0.002) node {H};

\end{axis}
\end{tikzpicture}

%% file: LiH-PNOF5-ERPA0.tex
\begin{tikzpicture}
\begin{axis}[
    width=0.45\textwidth,
    enlargelimits=false,
    cycle list name = matplotlib,
    xmin = 1.4,
    ymax = -7.7,
    ymin = -8.06,
    ytick={-8.06,-8.00,...,},
    xlabel = r (Bohr),
    ylabel = E (Ha),
    y tick label style={
        /pgf/number format/.cd,
        fixed,
        fixed zerofill,
        precision=2,
        /tikz/.cd
    },
    every axis y label/.style={
    at={(ticklabel* cs:1.05)},
    anchor=south,
    },
    legend style={/tikz/every even column/.append style={column sep=0.5cm},at={(0.77,0.2)},anchor=north,font=\tiny},
    ylabel style={align=center, inner sep=0pt, font=\footnotesize\bfseries\boldmath},
    xlabel style={align=center, inner sep=0pt, font=\footnotesize\bfseries\boldmath},
    x tick label style={font=\scriptsize\bfseries\boldmath},
    y tick label style={font=\scriptsize\bfseries\boldmath},
]

\addplot+[steelblue31119180,forget plot] table[x=r, y=E0] {lih-fci.dat};
\addplot+[darkorange25512714,forget plot] table[x=r, y=E1] {lih-fci.dat};
\addplot+[crimson2143940,forget plot] table[x=r, y=E3] {lih-fci.dat};
\addplot+[mediumpurple148103189,forget plot] table[x=r, y=E4] {lih-fci.dat};
\addplot+[sienna1408675,forget plot] table[x=r, y=E5] {lih-fci.dat};
\addplot+[orchid227119194,forget plot] table[x=r, y=E6] {lih-fci.dat};
\addplot+[gray127,forget plot] table[x=r, y=E7] {lih-fci.dat};
\addplot+[goldenrod18818934,forget plot] table[x=r, y=E8] {lih-fci.dat};

\addplot[] coordinates {(1,1)(2,2)};
\addlegendentry{FCI}

\addplot+[steelblue31119180,only marks,mark size=1pt,forget plot] table[x=r, y=E0] {lih-pnof5-erpa0.dat};
\addplot+[darkorange25512714,only marks,mark size=1pt,forget plot] table[x=r, y=E1] {lih-pnof5-erpa0.dat};
\addplot+[crimson2143940,only marks,mark size=1pt,forget plot] table[x=r, y=E3] {lih-pnof5-erpa0.dat};
\addplot+[mediumpurple148103189,only marks,mark size=1pt,forget plot] table[x=r, y=E4] {lih-pnof5-erpa0.dat};
\addplot+[sienna1408675,only marks,mark size=1pt,forget plot] table[x=r, y=E5] {lih-pnof5-erpa0.dat};
\addplot+[orchid227119194,only marks,mark size=1pt,forget plot] table[x=r, y=E6] {lih-pnof5-erpa0.dat};
\addplot+[gray127,only marks,mark size=1pt,forget plot] table[x=r, y=E7] {lih-pnof5-erpa0.dat};
\addplot+[goldenrod18818934,only marks,mark size=1pt,forget plot] table[x=r, y=E8] {lih-pnof5-erpa0.dat};

\addplot[only marks,mark size=2pt] coordinates {(1,1)(2,2)};
\addlegendentry{PNOF5-ERPA0}

\draw[fill=yellow!85!black,text=white] (2.95,-7.98) ellipse (0.45 and 0.023) node {Li};
\draw (3.40,-7.98) -- (3.65,-7.98);
\draw[fill=gray!100,text=white] (4.0,-7.98) ellipse (0.35 and 0.017) node {H};

\node[black!70] at (8.6,-7.765) {\tiny Li($^2$S) + H(1s)};
\node[black!70] at (8.6,-7.876) {\tiny Li($^2$P) + H(1s)};
\node[black!70] at (8.6,-7.95) {\tiny Li($^2$S) + H(1s)};

\node[steelblue31119180] at (3.0,-8.03) {\tiny $^1\Sigma^{+}$};
\node[darkorange25512714] at (6.0,-7.917) {\tiny $^1\Sigma^{+}$};
\node[crimson2143940] at (4.0,-7.875) {\tiny $^1\Pi$};
\node[mediumpurple148103189] at (6.25,-7.85) {\tiny $^1\Sigma^{+}$};
\node[gray127] at (3.5,-7.78) {\tiny $^1\Sigma^{+}$};
\node[goldenrod18818934] at (6.5,-7.78) {\tiny $^1\Sigma^{+}$};

\end{axis}
\end{tikzpicture}

%% file: LiH-PNOF5-ERPA.tex
\begin{tikzpicture}
\begin{axis}[
    width=\textwidth,
    enlargelimits=false,
    cycle list name = matplotlib,
    xmin = 1.4,
    ymax = -7.7,
    ymin = -7.95,
    xlabel = r (Bohr),
    ylabel = E (Ha),
    y tick label style={
        /pgf/number format/.cd,
        fixed,
        fixed zerofill,
        precision=2,
        /tikz/.cd
    },
    every axis y label/.style={
    at={(ticklabel* cs:1.05)},
    anchor=south,
    },
    legend style={/tikz/every even column/.append style={column sep=0.5cm},at={(0.37,0.45)},anchor=north,font=\tiny},
    ylabel style={align=center, inner sep=0pt, font=\footnotesize\bfseries\boldmath},
    xlabel style={align=center, inner sep=0pt, font=\footnotesize\bfseries\boldmath},
    x tick label style={font=\scriptsize\bfseries\boldmath},
    y tick label style={font=\scriptsize\bfseries\boldmath},
]

\addplot+[darkorange25512714,forget plot] table[x=r, y=E1] {lih-fci.dat};
\addplot+[mediumpurple148103189,forget plot] table[x=r, y=E4] {lih-fci.dat};
\addplot+[gray127,forget plot] table[x=r, y=E7] {lih-fci.dat};
\addplot+[goldenrod18818934,forget plot] table[x=r, y=E8] {lih-fci.dat};
\addplot[] coordinates {(1,1)(2,2)};
\addlegendentry{FCI}

\addplot+[darkorange25512714!40,only marks,mark size=1pt,mark=triangle*,forget plot] table[x=r, y=E1] {lih-pnof5-erpa0.dat};
\addplot+[mediumpurple148103189!40,only marks,mark size=1pt,mark=triangle*,forget plot] table[x=r, y=E4] {lih-pnof5-erpa0.dat};
\addplot+[gray127!40,only marks,mark size=1pt,mark=triangle*,forget plot] table[x=r, y=E7] {lih-pnof5-erpa0.dat};
\addplot+[goldenrod18818934!40,only marks,mark size=1pt,mark=triangle*,forget plot] table[x=r, y=E8] {lih-pnof5-erpa0.dat};

\addplot[black!40,only marks,mark size=2pt,mark=triangle*] coordinates {(1,1)(2,2)};
\addlegendentry{PNOF5-ERPA0}

\addplot+[darkorange25512714,only marks,mark size=1pt,mark=diamond*,forget plot] table[x=r, y=E1] {lih-pnof5-erpa.dat};
\addplot+[mediumpurple148103189,only marks,mark size=1pt,mark=diamond*,forget plot] table[x=r, y=E4] {lih-pnof5-erpa.dat};
\addplot+[gray127,only marks,mark size=1pt,mark=diamond*,forget plot] table[x=r, y=E7] {lih-pnof5-erpa.dat};
\addplot+[goldenrod18818934,only marks,mark size=1pt,mark=diamond*,forget plot] table[x=r, y=E8] {lih-pnof5-erpa.dat};

\addplot[only marks,mark size=2pt,mark=diamond*] coordinates {(1,1)(2,2)};
\addlegendentry{PNOF5-ERPA1}

\draw[fill=yellow!83!black,text=white] (7.45,-7.875) ellipse (0.45 and 0.0165) node {Li};
\draw (7.90,-7.875) -- (8.15,-7.875);
\draw[fill=gray!100,text=white] (8.5,-7.875) ellipse (0.35 and 0.0128) node {H};

\node[darkorange25512714] at (6.0,-7.926) {\tiny $^1\Sigma^{+}$};
\node[mediumpurple148103189] at (7.5,-7.84) {\tiny $^1\Sigma^{+}$};
\node[gray127] at (3.5,-7.785) {\tiny $^1\Sigma^{+}$};
\node[goldenrod18818934] at (7,-7.769) {\tiny $^1\Sigma^{+}$};

\end{axis}
\end{tikzpicture}

%% file: LiH-PNOF5-ERPA2.tex
\begin{tikzpicture}
\begin{axis}[
    width=\textwidth,
    enlargelimits=false,
    cycle list name = matplotlib,
    xmin = 1.4,
    ymax = -7.7,
    ymin = -7.95,
    xlabel = r (Bohr),
    ylabel = E (Ha),
    y tick label style={
        /pgf/number format/.cd,
        fixed,
        fixed zerofill,
        precision=2,
        /tikz/.cd
    },
    every axis y label/.style={
    at={(ticklabel* cs:1.05)},
    anchor=south,
    },
    legend style={/tikz/every even column/.append style={column sep=0.5cm},at={(0.37,0.45)},anchor=north,font=\tiny},
    ylabel style={align=center, inner sep=0pt, font=\footnotesize\bfseries\boldmath},
    xlabel style={align=center, inner sep=0pt, font=\footnotesize\bfseries\boldmath},
    x tick label style={font=\scriptsize\bfseries\boldmath},
    y tick label style={font=\scriptsize\bfseries\boldmath},
]

\addplot+[darkorange25512714,forget plot] table[x=r, y=E1] {lih-fci.dat};
\addplot+[mediumpurple148103189,forget plot] table[x=r, y=E4] {lih-fci.dat};
\addplot+[gray127,forget plot] table[x=r, y=E7] {lih-fci.dat};
\addplot+[goldenrod18818934,forget plot] table[x=r, y=E8] {lih-fci.dat};

\addplot[] coordinates {(1,1)(2,2)};
\addlegendentry{FCI}

\addplot+[darkorange25512714!40,only marks,mark size=1pt,mark=triangle*,forget plot] table[x=r, y=E1] {lih-pnof5-erpa0.dat};
\addplot+[mediumpurple148103189!40,only marks,mark size=1pt,mark=triangle*,forget plot] table[x=r, y=E4] {lih-pnof5-erpa0.dat};
\addplot+[gray127!40,only marks,mark size=1pt,mark=triangle*,forget plot] table[x=r, y=E7] {lih-pnof5-erpa0.dat};
\addplot+[goldenrod18818934!40,only marks,mark size=1pt,mark=triangle*,forget plot] table[x=r, y=E8] {lih-pnof5-erpa0.dat};

\addplot[black!40,only marks,mark size=2pt,mark=triangle*] coordinates {(1,1)(2,2)};
\addlegendentry{PNOF5-ERPA0}

\addplot+[darkorange25512714,only marks,mark size=1pt,mark=pentagon*,forget plot] table[x=r, y=E1] {lih-pnof5-erpa2.dat};
\addplot+[mediumpurple148103189,only marks,mark size=1pt,mark=pentagon*,forget plot] table[x=r, y=E4] {lih-pnof5-erpa2.dat};
\addplot+[gray127,only marks,mark size=1pt,mark=pentagon*,forget plot] table[x=r, y=E7] {lih-pnof5-erpa2.dat};
\addplot+[goldenrod18818934,only marks,mark size=1pt,mark=pentagon*,forget plot] table[x=r, y=E8] {lih-pnof5-erpa2.dat};

\addplot[only marks,mark size=2pt,mark=pentagon*] coordinates {(1,1)(2,2)};
\addlegendentry{PNOF5-ERPA2}

\draw[fill=yellow!83!black,text=white] (7.45,-7.875) ellipse (0.45 and 0.0165) node {Li};
\draw (7.90,-7.875) -- (8.15,-7.875);
\draw[fill=gray!100,text=white] (8.5,-7.875) ellipse (0.35 and 0.0128) node {H};

\node[darkorange25512714] at (6.0,-7.917) {\tiny $^1\Sigma^{+}$};
\node[mediumpurple148103189] at (7.5,-7.84) {\tiny $^1\Sigma^{+}$};
\node[gray127] at (3.5,-7.785) {\tiny $^1\Sigma^{+}$};
\node[goldenrod18818934] at (7,-7.775) {\tiny $^1\Sigma^{+}$};

\end{axis}
\end{tikzpicture}

%% file: LiH-PNOF7-ERPA2.tex
\begin{tikzpicture}
\begin{axis}[
    width=\textwidth,
    enlargelimits=false,
    cycle list name = matplotlib,
    xmin = 1.4,
    ymax = -7.7,
    ymin = -8.1,
    xlabel = r (Bohr),
    ylabel = E (Ha),
    y tick label style={
        /pgf/number format/.cd,
        fixed,
        fixed zerofill,
        precision=1,
        /tikz/.cd
    },
    every axis y label/.style={
    at={(ticklabel* cs:1.05)},
    anchor=south,
    },
    legend style={/tikz/every even column/.append style={column sep=0.5cm},at={(0.77,0.25)},anchor=north,font=\tiny},
    ylabel style={align=center, inner sep=0pt, font=\footnotesize\bfseries\boldmath},
    xlabel style={align=center, inner sep=0pt, font=\footnotesize\bfseries\boldmath},
    x tick label style={font=\scriptsize\bfseries\boldmath},
    y tick label style={font=\scriptsize\bfseries\boldmath},
]

\addplot+[steelblue31119180,forget plot] table[x=r, y=E0] {lih-fci.dat};
\addplot+[darkorange25512714,forget plot] table[x=r, y=E1] {lih-fci.dat};
\addplot+[forestgreen4416044,forget plot] table[x=r, y=E2] {lih-fci.dat};
\addplot+[crimson2143940,forget plot] table[x=r, y=E3] {lih-fci.dat};
\addplot+[mediumpurple148103189,forget plot] table[x=r, y=E4] {lih-fci.dat};
\addplot+[sienna1408675,forget plot] table[x=r, y=E5] {lih-fci.dat};
\addplot+[orchid227119194,forget plot] table[x=r, y=E6] {lih-fci.dat};
\addplot+[gray127,forget plot] table[x=r, y=E7] {lih-fci.dat};
\addplot+[goldenrod18818934,forget plot] table[x=r, y=E8] {lih-fci.dat};

\addplot[] coordinates {(1,1)(2,2)};
\addlegendentry{FCI}

\addplot+[steelblue31119180!30,only marks,mark size=1pt,mark=triangle*,forget plot] table[x=r, y=E0] {lih-pnof5-erpa2.dat};
\addplot+[darkorange25512714!30,only marks,mark size=1pt,mark=triangle*,forget plot] table[x=r, y=E1] {lih-pnof5-erpa2.dat};
\addplot+[crimson2143940!30,only marks,mark size=1pt,mark=triangle*,forget plot] table[x=r, y=E3] {lih-pnof5-erpa2.dat};
\addplot+[mediumpurple148103189!30,only marks,mark size=1pt,mark=triangle*,forget plot] table[x=r, y=E4] {lih-pnof5-erpa2.dat};
\addplot+[orchid227119194!30,only marks,mark size=1pt,mark=triangle*,forget plot] table[x=r, y=E6] {lih-pnof5-erpa2.dat};
\addplot+[gray127!30,only marks,mark size=1pt,mark=triangle*,forget plot] table[x=r, y=E7] {lih-pnof5-erpa2.dat};
\addplot+[goldenrod18818934!30,only marks,mark size=1pt,mark=triangle*,forget plot] table[x=r, y=E8] {lih-pnof5-erpa2.dat};

\addplot[only marks,mark size=2pt,mark=triangle*,black!30] coordinates {(1,1)(2,2)};
\addlegendentry{PNOF5-ERPA2}

\addplot+[steelblue31119180,only marks,mark size=1pt,mark=diamond*,forget plot] table[x=r, y=E0] {lih-pnof7-erpa2.dat};
\addplot+[darkorange25512714,only marks,mark size=1pt,mark=diamond*,forget plot] table[x=r, y=E1] {lih-pnof7-erpa2.dat};
\addplot+[crimson2143940,only marks,mark size=1pt,mark=diamond*,forget plot] table[x=r, y=E3] {lih-pnof7-erpa2.dat};
\addplot+[mediumpurple148103189,only marks,mark size=1pt,mark=diamond*,forget plot] table[x=r, y=E4] {lih-pnof7-erpa2.dat};
\addplot+[orchid227119194,only marks,mark size=1pt,mark=diamond*,forget plot] table[x=r, y=E6] {lih-pnof7-erpa2.dat};
\addplot+[gray127,only marks,mark size=1pt,mark=diamond*,forget plot] table[x=r, y=E7] {lih-pnof7-erpa2.dat};
\addplot+[goldenrod18818934,only marks,mark size=1pt,mark=diamond*,forget plot] table[x=r, y=E8] {lih-pnof7-erpa2.dat};

\addplot[only marks,mark size=2pt,mark=diamond*] coordinates {(1,1)(2,2)};
\addlegendentry{PNOF7-ERPA2}

\draw[fill=yellow!83!black,text=white] (2.95,-7.98) ellipse (0.45 and 0.026) node {Li};
\draw (3.40,-7.98) -- (3.65,-7.98);
\draw[fill=gray!100,text=white] (4.0,-7.98) ellipse (0.35 and 0.018) node {H};

\node[steelblue31119180] at (3.0,-8.03) {\tiny $^1\Sigma^{+}$};
\node[darkorange25512714] at (6.0,-7.917) {\tiny $^1\Sigma^{+}$};
\node[crimson2143940] at (4.0,-7.875) {\tiny $^1\Pi$};
\node[mediumpurple148103189] at (7.0,-7.86) {\tiny $^1\Sigma^{+}$};
\node[gray127] at (3.5,-7.78) {\tiny $^1\Sigma^{+}$};
\node[goldenrod18818934] at (7,-7.77) {\tiny $^1\Sigma^{+}$};

\end{axis}
\end{tikzpicture}

%% file: LiH-GNOF-ERPA2.tex
\begin{tikzpicture}
\begin{axis}[
    width=\textwidth,
    enlargelimits=false,
    cycle list name = matplotlib,
    xmin = 1.4,
    ymax = -7.7,
    ymin = -8.1,
    xlabel = r (Bohr),
    ylabel = E (Ha),
    y tick label style={
        /pgf/number format/.cd,
        fixed,
        fixed zerofill,
        precision=1,
        /tikz/.cd
    },
    every axis y label/.style={
    at={(ticklabel* cs:1.05)},
    anchor=south,
    },
    legend style={/tikz/every even column/.append style={column sep=0.5cm},at={(0.77,0.25)},anchor=north,font=\tiny},
    ylabel style={align=center, inner sep=0pt, font=\footnotesize\bfseries\boldmath},
    xlabel style={align=center, inner sep=0pt, font=\footnotesize\bfseries\boldmath},
    x tick label style={font=\scriptsize\bfseries\boldmath},
    y tick label style={font=\scriptsize\bfseries\boldmath},
]

\addplot+[steelblue31119180,forget plot] table[x=r, y=E0] {lih-fci.dat};
\addplot+[darkorange25512714,forget plot] table[x=r, y=E1] {lih-fci.dat};
\addplot+[forestgreen4416044,forget plot] table[x=r, y=E2] {lih-fci.dat};
\addplot+[crimson2143940,forget plot] table[x=r, y=E3] {lih-fci.dat};
\addplot+[mediumpurple148103189,forget plot] table[x=r, y=E4] {lih-fci.dat};
\addplot+[sienna1408675,forget plot] table[x=r, y=E5] {lih-fci.dat};
\addplot+[orchid227119194,forget plot] table[x=r, y=E6] {lih-fci.dat};
\addplot+[gray127,forget plot] table[x=r, y=E7] {lih-fci.dat};
\addplot+[goldenrod18818934,forget plot] table[x=r, y=E8] {lih-fci.dat};

\addplot[] coordinates {(1,1)(2,2)};
\addlegendentry{FCI}

\addplot+[steelblue31119180!30,only marks,mark size=1pt,mark=triangle*,forget plot] table[x=r, y=E0] {lih-pnof5-erpa2.dat};
\addplot+[darkorange25512714!30,only marks,mark size=1pt,mark=triangle*,forget plot] table[x=r, y=E1] {lih-pnof5-erpa2.dat};
\addplot+[crimson2143940!30,only marks,mark size=1pt,mark=triangle*,forget plot] table[x=r, y=E3] {lih-pnof5-erpa2.dat};
\addplot+[mediumpurple148103189!30,only marks,mark size=1pt,mark=triangle*,forget plot] table[x=r, y=E4] {lih-pnof5-erpa2.dat};
\addplot+[orchid227119194!30,only marks,mark size=1pt,mark=triangle*,forget plot] table[x=r, y=E6] {lih-pnof5-erpa2.dat};
\addplot+[gray127!30,only marks,mark size=1pt,mark=triangle*,forget plot] table[x=r, y=E7] {lih-pnof5-erpa2.dat};
\addplot+[goldenrod18818934!30,only marks,mark size=1pt,mark=triangle*,forget plot] table[x=r, y=E8] {lih-pnof5-erpa2.dat};

\addplot[only marks,mark size=2pt,mark=triangle*,black!30] coordinates {(1,1)(2,2)};
\addlegendentry{PNOF5-ERPA2}

\addplot+[steelblue31119180,only marks,mark size=1pt,mark=pentagon*,forget plot] table[x=r, y=E0] {lih-gnof-erpa2.dat};
\addplot+[darkorange25512714,only marks,mark size=1pt,mark=pentagon*,forget plot] table[x=r, y=E1] {lih-gnof-erpa2.dat};
\addplot+[crimson2143940,only marks,mark size=1pt,mark=pentagon*,forget plot] table[x=r, y=E3] {lih-gnof-erpa2.dat};
\addplot+[mediumpurple148103189,only marks,mark size=1pt,mark=pentagon*,forget plot] table[x=r, y=E4] {lih-gnof-erpa2.dat};
\addplot+[orchid227119194,only marks,mark size=1pt,mark=pentagon*,forget plot] table[x=r, y=E6] {lih-gnof-erpa2.dat};
\addplot+[gray127,only marks,mark size=1pt,mark=pentagon*,forget plot] table[x=r, y=E7] {lih-gnof-erpa2.dat};
\addplot+[goldenrod18818934,only marks,mark size=1pt,mark=pentagon*,forget plot] table[x=r, y=E8] {lih-gnof-erpa2.dat};

\addplot[only marks,mark size=2pt,mark=pentagon*] coordinates {(1,1)(2,2)};
\addlegendentry{GNOF-ERPA2}

\draw[fill=yellow!83!black,text=white] (2.95,-7.98) ellipse (0.45 and 0.026) node {Li};
\draw (3.40,-7.98) -- (3.65,-7.98);
\draw[fill=gray!100,text=white] (4.0,-7.98) ellipse (0.35 and 0.018) node {H};

\node[black!70] at (8.6,-7.765) {\tiny Li($^2$S) + H(1s)};
\node[black!70] at (8.6,-7.876) {\tiny Li($^2$P) + H(1s)};
\node[black!70] at (8.6,-7.95) {\tiny Li($^2$S) + H(1s)};

\node[steelblue31119180] at (3.0,-8.03) {\tiny $^1\Sigma^{+}$};
\node[darkorange25512714] at (6.0,-7.917) {\tiny $^1\Sigma^{+}$};
\node[crimson2143940] at (4.0,-7.865) {\tiny $^1\Pi$};
\node[mediumpurple148103189] at (6.25,-7.85) {\tiny $^1\Sigma^{+}$};
\node[gray127] at (3.5,-7.77) {\tiny $^1\Sigma^{+}$};
\node[goldenrod18818934] at (6.5,-7.775) {\tiny $^1\Sigma^{+}$};

\end{axis}
\end{tikzpicture}

%% file: Li2-PNOF5-ERPA2.tex
\begin{tikzpicture}
\begin{axis}[
    width=\textwidth,
    enlargelimits=false,
    cycle list name = matplotlib,
    xmin = 1.4,
    ymax = -14.81,
    ymin = -14.97,
    ytick={-14.97,-14.93,...,},
    xlabel = r (Bohr),
    ylabel = E (Ha),
    y tick label style={
        /pgf/number format/.cd,
        fixed,
        fixed zerofill,
        precision=2,
        /tikz/.cd
    },
    every axis y label/.style={
    at={(ticklabel* cs:1.05)},
    anchor=south,
    },
    legend style={/tikz/every even column/.append style={column sep=0.5cm},at={(0.55,0.35)},anchor=north,font=\tiny},
    ylabel style={align=center, inner sep=0pt, font=\footnotesize\bfseries\boldmath},
    xlabel style={align=center, inner sep=0pt, font=\footnotesize\bfseries\boldmath},
    x tick label style={font=\scriptsize\bfseries\boldmath},
    y tick label style={font=\scriptsize\bfseries\boldmath},
]

\addplot+[steelblue31119180,forget plot] table[x=r, y=E0] {li2-fci.dat};
\addplot+[darkorange25512714,forget plot] table[x=r, y=E1] {li2-fci.dat};
\addplot+[crimson2143940,forget plot] table[x=r, y=E3] {li2-fci.dat};
\addplot+[mediumpurple148103189,forget plot] table[x=r, y=E4] {li2-fci.dat};
\addplot+[sienna1408675,forget plot] table[x=r, y=E5] {li2-fci.dat};
\addplot+[orchid227119194,forget plot] table[x=r, y=E6] {li2-fci.dat};
\addplot+[gray127,forget plot] table[x=r, y=E7] {li2-fci.dat};
\addplot+[goldenrod18818934,forget plot] table[x=r, y=E8] {li2-fci.dat};

\addplot[] coordinates {(1,1)(2,2)};
\addlegendentry{FCI}
\addplot+[steelblue31119180,only marks,mark size=1pt,mark=*,forget plot] table[x=r, y=E0] {li2-pnof5-erpa2.dat};
\addplot+[darkorange25512714,only marks,mark size=1pt,mark=*,forget plot] table[x=r, y=E1] {li2-pnof5-erpa2.dat};
\addplot+[crimson2143940,only marks,mark size=1pt,mark=*,forget plot] table[x=r, y=E3] {li2-pnof5-erpa2.dat};
\addplot+[mediumpurple148103189,only marks,mark size=1pt,mark=*,forget plot] table[x=r, y=E4] {li2-pnof5-erpa2.dat};
\addplot+[sienna1408675,only marks,mark size=1pt,mark=*,forget plot] table[x=r, y=E5] {li2-pnof5-erpa2.dat};
\addplot+[orchid227119194,only marks,mark size=1pt,mark=*,forget plot] table[x=r, y=E6] {li2-pnof5-erpa2.dat};
\addplot+[gray127,only marks,mark size=1pt,mark=*,forget plot] table[x=r, y=E7] {li2-pnof5-erpa2.dat};
\addplot+[goldenrod18818934,only marks,mark size=1pt,mark=*,forget plot] table[x=r, y=E8] {li2-pnof5-erpa2.dat};

\addplot[only marks,mark size=2pt] coordinates {(1,1)(2,2)};
\addlegendentry{PNOF5-ERPA2}

\draw[fill=yellow!83!black,text=white] (1.95,-14.95) ellipse (0.45 and 0.010) node {Li};
\draw (2.39,-14.95) -- (2.81,-14.95);
\draw[fill=yellow!83!black,text=white] (3.25,-14.95) ellipse (0.45 and 0.010) node {Li};

\node[black!70] at (8.6,-14.840) {\tiny Li($^2$P) + Li($^2$P)};
\node[black!70] at (8.6,-14.905) {\tiny Li($^2$P) + Li($^2$S)};
\node[black!70] at (8.6,-14.96) {\tiny Li($^2$S) + Li($^2$S)};

\node[steelblue31119180] at (5.0,-14.95) {\tiny $^1\Sigma^{+}_\text{g}$};
\node[darkorange25512714] at (5.5,-14.89) {\tiny $^1\Sigma^{+}_\text{u}$};
\node[crimson2143940] at (4.25,-14.860) {\tiny $^1\Pi_\text{u}$};
\node[mediumpurple148103189] at (5.0,-14.875) {\tiny $^1\Sigma^{+}_\text{g}$};
\node[sienna1408675] at (6.0,-14.855) {\tiny $^1\Pi_\text{g}$};
\node[gray127] at (9.5,-14.815) {\tiny $^1\Sigma^{+}_\text{u}$};
\node[goldenrod18818934] at (6.0,-14.837) {\tiny $^1\Sigma^{+}_\text{g}$};

\end{axis}
\end{tikzpicture}

%% file: Li2-GNOF-ERPA2.tex
\begin{tikzpicture}
\begin{axis}[
    width=\textwidth,
    enlargelimits=false,
    cycle list name = matplotlib,
    xmin = 1.4,
    ymax = -14.81,
    ymin = -14.97,
    ytick={-14.97,-14.93,...,},
    xlabel = r (Bohr),
    ylabel = E (Ha),
    y tick label style={
        /pgf/number format/.cd,
        fixed,
        fixed zerofill,
        precision=2,
        /tikz/.cd
    },
    every axis y label/.style={
    at={(ticklabel* cs:1.05)},
    anchor=south,
    },
    legend style={/tikz/every even column/.append style={column sep=0.5cm},at={(0.55,0.35)},anchor=north,font=\tiny},
    ylabel style={align=center, inner sep=0pt, font=\footnotesize\bfseries\boldmath},
    xlabel style={align=center, inner sep=0pt, font=\footnotesize\bfseries\boldmath},
    x tick label style={font=\scriptsize\bfseries\boldmath},
    y tick label style={font=\scriptsize\bfseries\boldmath},
]

\addplot+[steelblue31119180,forget plot] table[x=r, y=E0] {li2-fci.dat};
\addplot+[darkorange25512714,forget plot] table[x=r, y=E1] {li2-fci.dat};
\addplot+[crimson2143940,forget plot] table[x=r, y=E3] {li2-fci.dat};
\addplot+[mediumpurple148103189,forget plot] table[x=r, y=E4] {li2-fci.dat};
\addplot+[sienna1408675,forget plot] table[x=r, y=E5] {li2-fci.dat};
\addplot+[orchid227119194,forget plot] table[x=r, y=E6] {li2-fci.dat};
\addplot+[gray127,forget plot] table[x=r, y=E7] {li2-fci.dat};
\addplot+[goldenrod18818934,forget plot] table[x=r, y=E8] {li2-fci.dat};

\addplot[] coordinates {(1,1)(2,2)};
\addlegendentry{FCI}

\addplot+[steelblue31119180,only marks,mark size=1pt,mark=*,forget plot] table[x=r, y=E0] {li2-gnof-erpa2.dat};
\addplot+[darkorange25512714,only marks,mark size=1pt,mark=*,forget plot] table[x=r, y=E1] {li2-gnof-erpa2.dat};
\addplot+[crimson2143940,only marks,mark size=1pt,mark=*,forget plot] table[x=r, y=E3] {li2-gnof-erpa2.dat};
\addplot+[mediumpurple148103189,only marks,mark size=1pt,mark=*,forget plot] table[x=r, y=E4] {li2-gnof-erpa2.dat};
\addplot+[sienna1408675,only marks,mark size=1pt,mark=*,forget plot] table[x=r, y=E5] {li2-gnof-erpa2.dat};
\addplot+[orchid227119194,only marks,mark size=1pt,mark=*,forget plot] table[x=r, y=E6] {li2-gnof-erpa2.dat};
\addplot+[gray127,only marks,mark size=1pt,mark=*,forget plot] table[x=r, y=E7] {li2-gnof-erpa2.dat};
\addplot+[goldenrod18818934,only marks,mark size=1pt,mark=*,forget plot] table[x=r, y=E8] {li2-gnof-erpa2.dat};

\addplot[only marks,mark size=2pt] coordinates {(1,1)(2,2)};
\addlegendentry{GNOF-ERPA2}

\draw[fill=yellow!83!black,text=white] (1.95,-14.95) ellipse (0.45 and 0.010) node {Li};
\draw (2.39,-14.95) -- (2.81,-14.95);
\draw[fill=yellow!83!black,text=white] (3.25,-14.95) ellipse (0.45 and 0.010) node {Li};

\node[black!70] at (8.6,-14.840) {\tiny Li($^2$P) + Li($^2$P)};
\node[black!70] at (8.6,-14.905) {\tiny Li($^2$P) + Li($^2$S)};
\node[black!70] at (8.6,-14.96) {\tiny Li($^2$S) + Li($^2$S)};

\node[steelblue31119180] at (5.0,-14.95) {\tiny $^1\Sigma^{+}_\text{g}$};
\node[darkorange25512714] at (5.5,-14.89) {\tiny $^1\Sigma^{+}_\text{u}$};
\node[crimson2143940] at (4.25,-14.860) {\tiny $^1\Pi_\text{u}$};
\node[mediumpurple148103189] at (5.0,-14.875) {\tiny $^1\Sigma^{+}_\text{g}$};
\node[sienna1408675] at (6.0,-14.857) {\tiny $^1\Pi_\text{g}$};
\node[gray127] at (9.5,-14.822) {\tiny $^1\Sigma^{+}_\text{u}$};
\node[goldenrod18818934] at (6.0,-14.837) {\tiny $^1\Sigma^{+}_\text{g}$};

\end{axis}
\end{tikzpicture}

%% file: Li2-References.tex
\begin{tikzpicture}
\begin{axis}[
    width=0.45\textwidth,
    enlargelimits=false,
    cycle list name = matplotlib,
    legend pos=outer north east,
    xmin = 1.4,
    ymax = -14.78,
    ymin = -14.98,
    ytick={-14.98,-14.94,...,},
    xlabel = r (Bohr),
    ylabel = E (Ha),
    y tick label style={
        /pgf/number format/.cd,
        fixed,
        fixed zerofill,
        precision=2,
        /tikz/.cd
    },
    every axis y label/.style={
    at={(ticklabel* cs:1.05)},
    anchor=south,
    },
    legend style={/tikz/every even column/.append style={column sep=0.1cm},at={(0.55,1.125)},anchor=north,font=\tiny},
    legend columns=3,
    ylabel style={align=center, inner sep=0pt, font=\footnotesize\bfseries\boldmath},
    xlabel style={align=center, inner sep=0pt, font=\footnotesize\bfseries\boldmath},
    x tick label style={font=\scriptsize\bfseries\boldmath},
    y tick label style={font=\scriptsize\bfseries\boldmath},
]

\addplot+[black,forget plot] table[x=r, y=E0] {li2-fci.dat};
\addplot+[black,forget plot] table[x=r, y=E1] {li2-fci.dat};
\addplot+[black,forget plot] table[x=r, y=E2] {li2-fci.dat};
\addplot+[black,forget plot] table[x=r, y=E3] {li2-fci.dat};
\addplot+[black,forget plot] table[x=r, y=E4] {li2-fci.dat};
\addplot+[black,forget plot] table[x=r, y=E5] {li2-fci.dat};
\addplot+[black,forget plot] table[x=r, y=E6] {li2-fci.dat};
\addplot+[black,forget plot] table[x=r, y=E7] {li2-fci.dat};
\addplot+[black,forget plot] table[x=r, y=E8] {li2-fci.dat};

\addplot[] coordinates {(1,1)(2,2)};
\addlegendentry{FCI}

\addplot+[darkorange25512714!60,only marks,mark size=1pt,mark=*,forget plot] table[x=r, y=E0] {li2-cam-b3lyp.dat};
\addplot+[darkorange25512714!60,only marks,mark size=1pt,mark=*,forget plot] table[x=r, y=E1] {li2-cam-b3lyp.dat};
\addplot+[darkorange25512714!60,only marks,mark size=1pt,mark=*,forget plot] table[x=r, y=E2] {li2-cam-b3lyp.dat};
\addplot+[darkorange25512714!60,only marks,mark size=1pt,mark=*,forget plot] table[x=r, y=E3] {li2-cam-b3lyp.dat};
\addplot+[darkorange25512714!60,only marks,mark size=1pt,mark=*,forget plot] table[x=r, y=E4] {li2-cam-b3lyp.dat};
\addplot+[darkorange25512714!60,only marks,mark size=1pt,mark=*,forget plot] table[x=r, y=E5] {li2-cam-b3lyp.dat};
\addplot+[darkorange25512714!60,only marks,mark size=1pt,mark=*,forget plot] table[x=r, y=E6] {li2-cam-b3lyp.dat};
\addplot+[darkorange25512714!60,only marks,mark size=1pt,mark=*,forget plot] table[x=r, y=E7] {li2-cam-b3lyp.dat};
\addplot+[darkorange25512714!60,only marks,mark size=1pt,mark=*,forget plot] table[x=r, y=E8] {li2-cam-b3lyp.dat};

\addplot[darkorange25512714!60,only marks,mark size=2pt] coordinates {(1,1)(2,2)};
\addlegendentry{TD-CAM-B3LYP}

\addplot+[steelblue31119180,only marks,mark size=1pt,mark=*,forget plot] table[x=r, y=E0] {li2-gnof-erpa2.dat};
\addplot+[steelblue31119180,only marks,mark size=1pt,mark=*,forget plot] table[x=r, y=E1] {li2-gnof-erpa2.dat};
\addplot+[steelblue31119180,only marks,mark size=1pt,mark=*,forget plot] table[x=r, y=E2] {li2-gnof-erpa2.dat};
\addplot+[steelblue31119180,only marks,mark size=1pt,mark=*,forget plot] table[x=r, y=E4] {li2-gnof-erpa2.dat};
\addplot+[steelblue31119180,only marks,mark size=1pt,mark=*,forget plot] table[x=r, y=E5] {li2-gnof-erpa2.dat};
\addplot+[steelblue31119180,only marks,mark size=1pt,mark=*,forget plot] table[x=r, y=E6] {li2-gnof-erpa2.dat};
\addplot+[steelblue31119180,only marks,mark size=1pt,mark=*,forget plot] table[x=r, y=E7] {li2-gnof-erpa2.dat};
\addplot+[steelblue31119180,only marks,mark size=1pt,mark=*,forget plot] table[x=r, y=E8] {li2-gnof-erpa2.dat};

\addplot[steelblue31119180,only marks,mark size=2pt] coordinates {(1,1)(2,2)};
\addlegendentry{GNOF-ERPA2}


\draw[fill=yellow!85!black,text=white] (1.95,-14.96) ellipse (0.45 and 0.012) node {Li};
\draw (2.39,-14.96) -- (2.81,-14.96);
\draw[fill=yellow!85!black,text=white] (3.25,-14.96) ellipse (0.45 and 0.012) node {Li};

\end{axis}
\end{tikzpicture}

%% file: N2.tex
\begin{tikzpicture}   

\begin{groupplot}[
    group style={
        group name=my fancy plots,
        group size=2 by 3,
        xticklabels at=edge bottom,
        horizontal sep=4pt
    },
    ]

    \nextgroupplot[
    xmin=1.4,
    xmax=4.4,
    ymax = -108.3,
    ymin = -109.3,
    cycle list name = matplotlib,
    ylabel = E (Ha),
    ytick={-109.3,-109.1,...,},
    y tick label style={
        /pgf/number format/.cd,
        fixed,
        fixed zerofill,
        precision=1,
        /tikz/.cd
    },
    every axis y label/.style={
    at={(ticklabel* cs:1.05)},
    anchor=south,
    },
    legend style={/tikz/every even column/.append style={column sep=0.5cm},at={(0.43,0.9)},anchor=north,font=\tiny},
    ylabel style={align=center, inner sep=0pt, font=\footnotesize\bfseries\boldmath},
    xlabel style={align=center, inner sep=0pt, font=\footnotesize\bfseries\boldmath},
    x tick label style={font=\scriptsize\bfseries\boldmath},
    y tick label style={font=\scriptsize\bfseries\boldmath},
    width=0.36\textwidth,
    height=0.3\textwidth,
    ]
    
    \addplot+[steelblue31119180,forget plot] table[x=r, y=E0] {n2-cisdtq.dat};
    \addplot+[darkorange25512714,forget plot] table[x=r, y=E1] {n2-cisdtq.dat};

    \addplot[] coordinates {(2,-107)(3,-107)};
    \addlegendentry{CISDTQ}

    \addplot+[steelblue31119180,only marks,mark size=1pt,forget plot] table[x=r, y=E0] {n2-gnof-erpa0.dat};
    \addplot+[darkorange25512714,only marks,mark size=1pt,forget plot] table[x=r, y=E1] {n2-gnof-erpa0.dat};

    \addplot[only marks,mark size=2pt] coordinates {(2,-107)(3,-107)};
    \addlegendentry{GNOF-ERPA0}


    \draw (3.0,-109.18) -- (3.8,-109.18);
    \draw (3.0,-109.20) -- (3.8,-109.20);
    \draw (3.0,-109.22) -- (3.8,-109.22);
    \draw[fill=Green,text=white] (3.0,-109.2) ellipse (0.21 and 0.08) node {N};
    \draw[fill=Green,text=white] (3.8,-109.2) ellipse (0.21 and 0.08) node {N};

    \node[darkorange25512714] at (2.4,-108.87) {\tiny $^1\Sigma^{+}_\text{u}$};
    \node[steelblue31119180] at (2.1,-109.20) {\tiny $^1\Sigma^{+}_\text{g}$};

    \nextgroupplot[
    xmin=13.0,
    xmax=14.0,
    ymax = -108.3,
    ymin = -109.3,
    cycle list name = matplotlib,    
    ytick=\empty,
    ytick style={draw=none},
    axis x discontinuity=crunch,
    xtick={13,14},
    xlabel style={align=center, inner sep=0pt, font=\footnotesize\bfseries\boldmath},
    x tick label style={font=\scriptsize\bfseries\boldmath},
    width=0.20\textwidth,
    height=0.3\textwidth,
    ]
    \addplot+[steelblue31119180,forget plot] table[x=r, y=E0] {n2-cisdtq.dat};
    \addplot+[darkorange25512714,forget plot] table[x=r, y=E1] {n2-cisdtq.dat};

    \addplot+[steelblue31119180,only marks,mark size=1pt,forget plot] table[x=r, y=E0] {n2-gnof-erpa0.dat};
    \addplot+[darkorange25512714,only marks,mark size=1pt,forget plot] table[x=r, y=E1] {n2-gnof-erpa0.dat};

    \draw[dashed] (1,-108.76) -- (15,-108.76);
    \draw[dashed] (1,-108.96) -- (15,-108.96);

    \node[black!70] at (13.5,-108.725) {\tiny N($^2$D) + N($^2$D)};
    \node[black!70] at (13.5,-108.92) {\tiny N($^4$S) + N($^4$S)};        

    \nextgroupplot[
    xmin=1.4,
    xmax=4.4,
    ymax = 2.0,
    ymin = 1.0,
    cycle list name = matplotlib,
    ylabel = 2n$_\text{LSDONO}$ and 2n$_\text{LSDONO-1}$ (G.S.),
    ytick={0.0,0.50,1.00,1.50,2.00},
    y tick label style={
        /pgf/number format/.cd,
        fixed,
        fixed zerofill,
        precision=2,
        /tikz/.cd
    },
    every axis y label/.style={
    at={(ticklabel* cs:1.20)},
    xshift = 62,
    anchor=south,
    },
    ylabel style={align=center, inner sep=0pt, font=\footnotesize\bfseries\boldmath},
    xlabel style={align=center, inner sep=0pt, font=\footnotesize\bfseries\boldmath},
    x tick label style={font=\scriptsize\bfseries\boldmath},
    y tick label style={font=\scriptsize\bfseries\boldmath},
    width=0.36\textwidth,
    height=0.18\textwidth,
    ]
    \addplot+[crimson2143940,only marks,mark=*,mark size=1pt,forget plot] table[x=r, y=22] {GNOF-ns.dat};
    
    \nextgroupplot[
    xmin=13.0,
    xmax=14.0,
    ymax = 2.0,
    ymin = 1.0,
    cycle list name = matplotlib,    
    ytick=\empty,
    ytick style={draw=none},
    axis x discontinuity=crunch,
    xtick={13,14},
    xlabel style={align=center, inner sep=0pt, font=\footnotesize\bfseries\boldmath},
    x tick label style={font=\scriptsize\bfseries\boldmath},
    width=0.20\textwidth,
    height=0.18\textwidth,
    ]
    \addplot+[crimson2143940,only marks,mark=*,mark size=1pt,forget plot] table[x=r, y=22] {GNOF-ns.dat};

    \nextgroupplot[
    xmin=1.4,
    xmax=4.4,
    ymax = 2.0,
    ymin = 1.0,
    cycle list name = matplotlib,
    xlabel = r (Bohr),
    ylabel = 2n$_\text{LSDONO-2}$ (G.S.),
    ytick={0.0,0.50,1.00,1.50,2.00},
    y tick label style={
        /pgf/number format/.cd,
        fixed,
        fixed zerofill,
        precision=2,
        /tikz/.cd
    },
    every axis y label/.style={
    at={(ticklabel* cs:1.20)},
    xshift = 26,
    anchor=south,
    },
    ylabel style={align=center, inner sep=0pt, font=\footnotesize\bfseries\boldmath},
    xlabel style={align=center, inner sep=0pt, font=\footnotesize\bfseries\boldmath},
    x tick label style={font=\scriptsize\bfseries\boldmath},
    y tick label style={font=\scriptsize\bfseries\boldmath},
    width=0.36\textwidth,
    height=0.18\textwidth,
    ]
    \addplot+[forestgreen4416044,only marks,mark=*,mark size=1pt,forget plot] table[x=r, y=23] {GNOF-ns.dat};
    
    \nextgroupplot[
    xmin=13.0,
    xmax=14.0,
    ymax = 2.0,
    ymin = 1.0,
    cycle list name = matplotlib,    
    ytick=\empty,
    ytick style={draw=none},
    axis x discontinuity=crunch,
    xtick={13,14},
    xlabel style={align=center, inner sep=0pt, font=\footnotesize\bfseries\boldmath},
    x tick label style={font=\scriptsize\bfseries\boldmath},
    width=0.20\textwidth,
    height=0.18\textwidth,
    ]
    \addplot+[forestgreen4416044,only marks,mark=*,mark size=1pt,forget plot] table[x=r, y=23] {GNOF-ns.dat};
    
    \end{groupplot}
    
    \end{tikzpicture}